\documentclass{article}

\usepackage[total={6.5in, 9in}]{geometry}
\usepackage{graphicx}
\usepackage[caption = false]{subfig}
\usepackage[utf8]{inputenc} 
\usepackage[T1]{fontenc}    
\usepackage{hyperref}       
\usepackage{url}            
\usepackage{booktabs}       
\usepackage{amsmath}
\usepackage{amsfonts}       
\usepackage{mathtools}
\usepackage{bm}             
\usepackage{nicefrac}       
\usepackage{microtype}      
\usepackage{authblk}        

\DeclareMathOperator*{\argmax}{arg\,max}

\renewcommand{\eqref}[1]{Equation\,(\ref{#1})}
\newcommand{\figref}[1]{Figure\,\ref{#1}}

\begin{document}
\title{Decentralized Algorithms for \\ Consensus-Based Power Packet Distribution}
\author[1]{Seongcheol Baek}
\author[2]{Hiroyasu Ando}
\author[3]{Takashi Hikihara}
\affil[1,3]{Department of Electrical Engineering, Kyoto University}
\affil[2]{Faculty of Engineering, Information and Systems, University of Tsukuba}
{
    \makeatletter
    \renewcommand\AB@affilsepx{: \protect\Affilfont}
    \makeatother

    \affil[ ]{E-mail}

    \makeatletter
    \renewcommand\AB@affilsepx{, \protect\Affilfont}
    \makeatother

    \affil[1]{s-baek@dove.kuee.kyoto-u.ac.jp}
    \affil[2]{ando@sk.tsukuba.ac.jp}
    \affil[3]{hikihara.takashi.2n@kyoto-u.ac.jp}
}
\date{\thanks{This paper was submitted to Nonlinear Theory and Its Applications, IEICE on October 29, 2020}} 
\maketitle
\begin{abstract}
Power packets are proposed as a transmission unit 
that can deliver power and information simultaneously.
They are transferred using the store-and-forward method of power routers. 
A system that achieves power supply/demand in this manner
is called a power packet network (PPN).
A PPN is expected to enhance structural robustness and operational reliability
in an energy storage system (ESS) with recent diverse distributed sources. 
However, 
this technology is still in its early stage,
and faces numerous challenges,
such as high cost of implementation and complicated energy management.
In this paper,
we propose a novel power control based on decentralized algorithms for a PPN.
Specifically, 
the power supply is triggered and managed by communications between power routers.
We also discuss the mechanism of the decentralized algorithm for the operation of power packets
and reveal the feasibility of the given control method 
and application by forming biased power flows on the consensus-based distribution.
\end{abstract}



\section{Introduction}
With the recent emergence of distributed power sources, 
and the shifting trend from fuel engines to electrical motors along with the advancement of battery efficiency,
the decentralization technique has attracted considerable attention 
from the fields of mechanical, control, and network engineering.
The characteristics of the next-generation power system,
such as high fluctuations, in-demand generation, and combined use of batteries and other power sources,
have resulted in the consideration of more delicate controls
by system engineers. 
In a smart grid,
for example,
load peak shaving and power smoothing are the common methods
that are considered \cite{load_peak_shaving_distribution_grid,smart_grid_storage_controller}.
From the viewpoint of system design,
decentralization can be sufficiently applied to achieve scalability and flexibility,
thanks to the system operation based on the agent-to-agent communication and interaction \cite{decentralized_sensor_network}.

A power packet network (PPN) 
is motivated by an initial concept of open electric energy network (OEEN)
which is introduced to enable power trading between participants, 
such as in the energy industry in the USA \cite{open_network}.
Although the scale of distribution and the expected applications are different from its origin,
a PPN employs a key concept: the introduction of a communication system to power system \cite{power_processing}.
Specifically, 
in a PPN, 
the power routers generate power packets with pulsed power sequences,
and the semantic functions are granted by dividing the bit strings in the unit power packet
into informational tag and power payload \cite{power_processing,power_processing_advanced}.
Power packet transmissions between routers
are performed using the store-and-forward method \cite{router_power,exp_packet_generation}.
This method relaxes the strict balancing rule of supply/demand
and provides network buffer to shift power flow temporally and spatially.
Thus, 
a PPN is expected to exhibit high controllability in energy management 
as well as adaptivity to deal with diverse energy conditions.

Recent studies on PPN cover a variety of spectra,
such as dynamics, devices, control theory, energy storage system, and applications.
Nawata {\it et al.} proposed a symbol-based transmission model
that can explain power flow based on a symbol propagation matrix \cite{packet_transferability}.
T. Hikihara {\it et al.} proposed another theoretical approach
which suggested that the flow of power packets can be quantized and developed into the Schrodinger equation \cite{packet_dispatching_dynamics}.
In the related studies to PPN, 
the research group of E. Gelenbe has proposed a methodology 
for the implementation, utility, and optimization of the energy packet network
through the problem of energy packet distribution
on overlapped network of information processing layer and energy processing layer \cite{optimization_energy_packets,energy_packet_cloud}.
Moreover, 
related technical discussions have indicated the application of synchronization to enable a stable exchange of information between routers  \cite{packet_synchronization},
as well as a security strategy with the use of power packets \cite{security_power_packet}.
In relation to our previous works,
H. Ando {\it et al.} proposed the consensus dynamics to analyze the dynamics of the power packet transmission
and to capture characteristics originating from the connections on the network \cite{consensus_packet,bio_power_consensus}.
Moreover, 
we further discussed the relationship 
between the above model and the emulation initiated by the decentralized control of a PPN \cite{consensus_based_dc,aip_baek}.
This work provides a theoretical understanding of the consensus-based packet distribution
and the possibility of the decentralized control of a PPN.

With the recent advancements of distributed power sources,
scalability and flexibility have become important characteristics in the network design and its application.
A potential vision of a PPN has been proposed through the operation of motors and load control based on power packets \cite{close_angle_control,networked_packet_dispatching}.
However, 
a system with different purposes and structures used in the semantic/schematic design is costly.
Based on this context, 
a decentralized algorithm-based control that is independent of the topological structure of a PPN
is proposed,
thus eventually improving scalability, flexibility, and redundancy.
For power packet transmission, 
we employ the consensus-based distribution model 
and consider two operational strategies:
the top-down (or supply-driven) method and the bottom-up (or demand-driven) method.
Moreover, 
we reveal a feasible application of energy management 
using both strategies,
in which power flows can be prioritized to power routers of the bottom-up method.

The remainder of this paper is organized as follows.
Section \ref{sec:ppn} introduces the PPN.
Section \ref{sec:modeling} discusses the graphical analysis and interpretation of a PPN,
and the consensus-based distribution model to capture the dynamics of power flow.
Next, 
we propose a decentralized algorithm for a PPN
on the basis of the routers' communication and distribution model.
Section \ref{sec:sim} explains the packet distribution scenario for two network structures
and simulates the decentralized control of a PPN.
The results the simulation reveal the feasibility of the proposed control methods 
and the possibility for a priority-based power control using the top-down and bottom-up methods.
Finally, 
Section \ref{sec:conclusions} concludes the paper.

\section{Power Packet Network} \label{sec:ppn}
\begin{figure}[t!]
  \centering
  \includegraphics[width=0.95\textwidth]{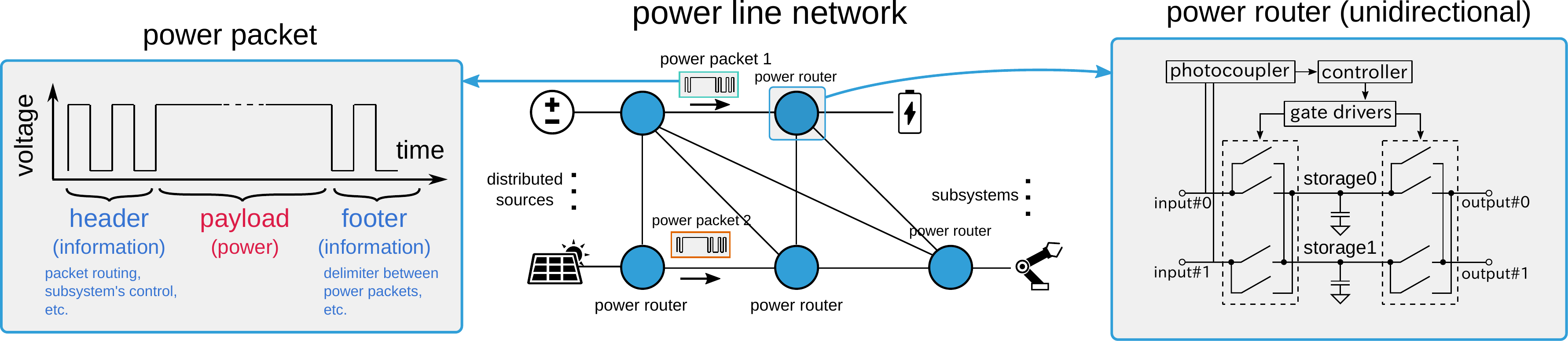}
  \caption{Layout of a power packet network (PPN). 
  Using the store-and-forward method, 
  the power routers deliver the power packets from the supply side to the demand side.}
  \label{f:ppdn}
\end{figure}

In \figref{f:ppdn},
the schematic diagram of a PPN is presented.
Given the pulsed direct current (DC) power, 
a power packet consists of bit sequences of information tag and power payload.
An information tag denoted as ``header'' and ``footer'', 
delivers information using coded bit sequences. 
Specifically, 
information on the routing address or control signal is placed in the ``header,'' 
and a delimiter is embedded in ``footer'' to signal the termination of packet transmission. 
The power payload physically delivers the electric power from the sender to the receiver.

A power router generates or receives the power packets with built-in switches and storages \cite{power_processing_advanced}.
This allows the group of routers to actively control the power flows on each of the connected ports. 
Specifically, 
a power router detects the inflow of power packets through a photocoupler. 
The incoming header information is processed in the controller 
and delivered to the gate driver to control each of the built-in switches, 
thus opening the corresponding port to charge built-in storages with power inflow (payload). 
Subsequently, 
if the footer is detected, 
the online port is closed, 
and the transmission of the given power packet is terminated.
In addition to storing power packets, 
a power router can also forward power packets by generating pulses with switches and charge storage. 
The use of the store-and-forward method
results in the quantization of power flows and simultaneous transmission of both power and information.

With the implementation of power routers, 
a PPN can accommodate multilevel DC sources 
and deliver power packets to the demanding subsystem.
Since power packets contain routing information and optionally control information,
the transmission network can handle several loads (subsystem) with shared lines,
which is called time-division multiplexing.
In terms of the recent progress of distributed power sources,
the integrity and flexibility of PPN are expected to solve the emerging issues on power management,
such as energy security, sporadic growth of grid, large loss due to oversupply, and unpredictability of renewable energy \cite{barrett2016challenges}.

The structure of a PPN lies in its switching topology,
which is a notable feature.
Specifically,
the paths where power packets go through are not constant.
Each path gets connected or disconnected, 
depending on the routers' operation \cite{consensus_cooperation}.
Another feature of a PPN is that the power packet transmission is controlled autonomously by routers,
which is thanks to the high-speed field-programmable gate array platform \cite{FPGA}.
In addition, 
the autonomous operation on each power router leads to 
the possibility of a decentralized control system (DCS) on PPN. 
A DCS improves the reliability in both energy management and failure management,
as well as structural scalability
that is a challenge in the previous power system. 

\section{Modeling of Decentralized Algorithm for a Power Packet Network} \label{sec:modeling}
Provided that a system is driven in a decentralized manner,
the spread dynamics of physical quantity or information
can be explained based on the agent-to-agent interaction.
Given this idea,
we employ the consensus dynamics to elucidate the distribution of power flows on a PPN and model a decentralized algorithm.

\subsection{Outline of the Graphical Analysis of a Power Packet Network} \label{subsec:graphical_analysis_ppn}
For simplicity, 
we assume that each of the routers is bidirectional and has a single storage or DC source.
Given a transmission network, 
such as in \figref{f:ppdn},
we consider a graph notation $G(\bm{V}, \bm{E}, \bm{w}, \bm{c})$, 
with $\bm{V}$ denoting the nodes; $\bm{E}$, the edges; $\bm{w}$, the node weight; and $\bm{c}$, the edge weight.
The nodes represent a set of power sources, a storage in routers, and outflow sinks (GNDs).
The edges are given as transmission paths, 
including the ideal line resistance.
The generalized impedance can also be considered in the analysis of the algebraic graph \cite{general_analytical_solutions,spectral_graph_applications}.
Thus, 
any edge between two connected nodes 
involves the line resistance $r$, 
of which reverse is given as edge weight $w$, {\it i.e.}, $w_{ij}=r_{ij}^{-1}$ for path $e_{ij}$.
The node weights are given as capacitances of nodes
indicating that the storage capacitance $c_i$ is directly applied as a node weight of node $u_{i}$.
In case of sources and sinks,
we assume that the node weight is a positive infinity.

\begin{figure}[t!]
  \centering
  \includegraphics[width=0.55\textwidth]{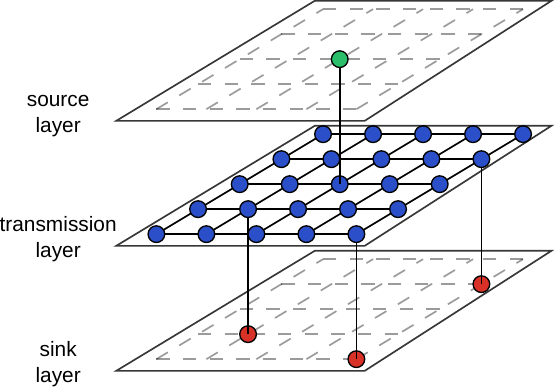}
  \caption{A three-layered schematic of a PPN.
  The nodes in the source layer, transmission layer, and sink layer 
  indicate the power routers with a source, power routers with a storage, and GNDs, respectively.
  The on-resistances of the routers' switching devices and loads are located on each path.
  The power flows through the source layer
  and is distributed to the transmission layer;
  it then flows out through the sink layer.
  }
  \label{f:multilevel_network}
\end{figure}

\figref{f:multilevel_network} presents a multilayered schematic network of $G(\bm{V}, \bm{E}, \bm{w}, \bm{c})$
based on the divided three node groups.
The divided layers facilitate the visualization of power flows between sources and sinks.
From the given schematic, 
the node voltage vector on the source layer is denoted as ${\bm v}_{\rm src}$, 
the node voltage vector on the transmission layer as ${\bm v}$, 
and the rest on the sink layer as ${\bm v}_{\rm snk}$.
Here, 
${\bm v}_{\rm src}$ denotes a positive constant vector,
and ${\bm v}_{\rm snk}$ denotes a zero constant vector.
\begin{align}
  {\bm v}_{\rm src} &= {\rm const.} > 0 \\
  {\bm v}_{\rm snk} &= {\rm 0}
\end{align}
Each of the paths between the nodes has a weight;
if path $e_{ij}$ is switchable, 
its weight is defined as $\{0, r_{ij}^{-1}\}$, 
depending on its switching state.
If the path is routed, 
the weight is given as $r_{ij}^{-1}$; 
otherwise, 
$0$.
This is reasonable because if the path is disconnected, 
the line resistance is obtained as $\infty$,
of which reverse is $0$.

For a multilayered network, 
the spread dynamics of matters or information is considered.
The consensus dynamics provides a useful analysis for problems,
such as collective behavior of flocks and swarms, 
formation control for multirobot systems, 
synchronization of coupled oscillators, 
consensus-based belief propagation, 
and so on \cite{consensus_cooperation,consensus_problems,qualitative_problems}.
Given a variable $x_{i\in\{0,\cdots, n-1\}}$ for each node on graph $G$,
the spread process is described as follows in terms of node $u_{i}$:
\begin{equation}
  c_i\dot{x}_i = \sum_{j\in\{k; (i,k)\in E(G)\}} w_{ij}(x_j - x_i) + b_i,
  \label{e:condyn_ct_agent}
\end{equation}
where $c_i$ indicates the node weight; $w_{ij}$, the edge weight; and $b_i$, bias.
Here, 
we assume a general variable $x_i$, 
such as matters or information in the system.
The above equation is generalized using the weighted Laplacian (or edge-weight Laplacian matrix) $\bm{L}(G)$
and a node-weight matrix $\bm{C}={\rm diag}(c_0, \cdots, c_{n-1})$ \cite{consensus_cooperation}.
\begin{equation}
  \bm{C}\dot{\bm{x}} = -\bm{L}\bm{x} + \bm{b}
  \label{e:condyn_ct}
\end{equation}
Following this, 
the relation based on a discrete time is obtained as follows.
\begin{equation}
  \bm{x}(k+1) = (\bm{I} - \epsilon\bm{C}^{-1}\bm{L})\bm{x}(k) + \epsilon\bm{C}^{-1}\bm{b}
  \label{e:condyn_dt}
\end{equation}
The state-transition matrix, 
which is expressed as $(\bm{I} - \epsilon\bm{C}^{-1}\bm{L})$ in \eqref{e:condyn_dt},
is stable in the case of connected graph $G$ and $0 < \epsilon < 1 / \max_{i\in V}(c_i^{-1} \sum_jw_{ij})$;
thus, with condition $\bm{b}=\bm{0}$, 
each element of $\bm{x}$ asymptotically reaches a weighted average.
The consensus dynamics further provides an estimation for the spread performance,
which is often based on algebraic connectivity $\lambda_2(\bm{L})$ or eigendecomposition \cite{consensus_cooperation,consensus_problems}.

\subsection{Power Dynamics of the Consensus-Based Distribution} \label{subsec: power dynamics}
On a PPN, 
such as in \figref{f:multilevel_network}, 
the occurrence of power packet transmissions can be assumed,
causing the power flows from the source layer to sink.
Let us consider a node voltage $\bm{v}$ for the variable $\bm{x}$ in \eqref{e:condyn_ct}.
Considering that $\bm{b}=0$ and the left term in \eqref{e:condyn_ct} is the net current of the node,
the following relation is derived:
\begin{equation}
  \bm{i}_{\rm net} = -\bm{L}\bm{v}
  \label{e:condyn_power}
\end{equation}
Now it can be easily observed that $\bm{L}$ denotes the inverse of network resistance,
because $\bm{w}=\{w_{ij}; e_{ij}=(i,j)\in \bm{E}\}$ is adopted based on the line resistance (or impedance generally).
Since the paths switch depending on the routers' operation,
the weighted Laplacian is a time-varying variable in terms of time $t$,
which is expressed as $\bm{L} = \bm{L}(S_t(G))$ with a subgraph $S_t(G) = G(\bm{V}, \bm{E}_t)$, 
where $\bm{E}_t \subset \bm{E}$.
From the three-layered model presented in \figref{f:multilevel_network}, 
\eqref{e:condyn_power} is further improved with current inflow $\bm{i}_{\rm in}$ and outflow $\bm{i}_{\rm out}$.
\begin{equation}
  \begin{bmatrix}
    \bm{i}_{\rm in} \\ \bm{C}\dot{\bm{v}} \\ \bm{i}_{\rm out}
  \end{bmatrix}
  =
  -
  \begin{bmatrix}
    {\bm L}_{11} & {\bm L}_{12} & {\bm L}_{13} \\
    {\bm L}_{21} & {\bm L}_{22} & {\bm L}_{23} \\
    {\bm L}_{31} & {\bm L}_{32} & {\bm L}_{33} \\
  \end{bmatrix}
  \begin{bmatrix}
    \bm{v}_{\rm src} \\ \bm{v} \\ \bm{v}_{\rm snk}
  \end{bmatrix},
  \label{e:i_ext=Lv2}
\end{equation}
Note that each element ${\bm L}_{ij}$ of the weighted Laplacian is time-varying. 
Assuming the constant voltages in the source and sink layers,
{\it i.e.}, $\bm{v}_{\rm src}= {\rm const.}$ and $\bm{v}_{\rm snk} = \bm{0}$,
the following dynamics can be obtained,
which explains the voltage distribution on a PPN.
\begin{equation}
  \dot{\bm{v}} = -\bm{C}^{-1}{\bm L}_{22}\bm{v} - \bm{C}^{-1}{\bm L}_{21}\bm{v}_{\rm src}
  \label{e:power_dynamics}
\end{equation}

\subsection{Modeling of Decentralized Algorithms}

\begin{figure}[t!]
  \centering
  \includegraphics[width=0.30\textwidth]{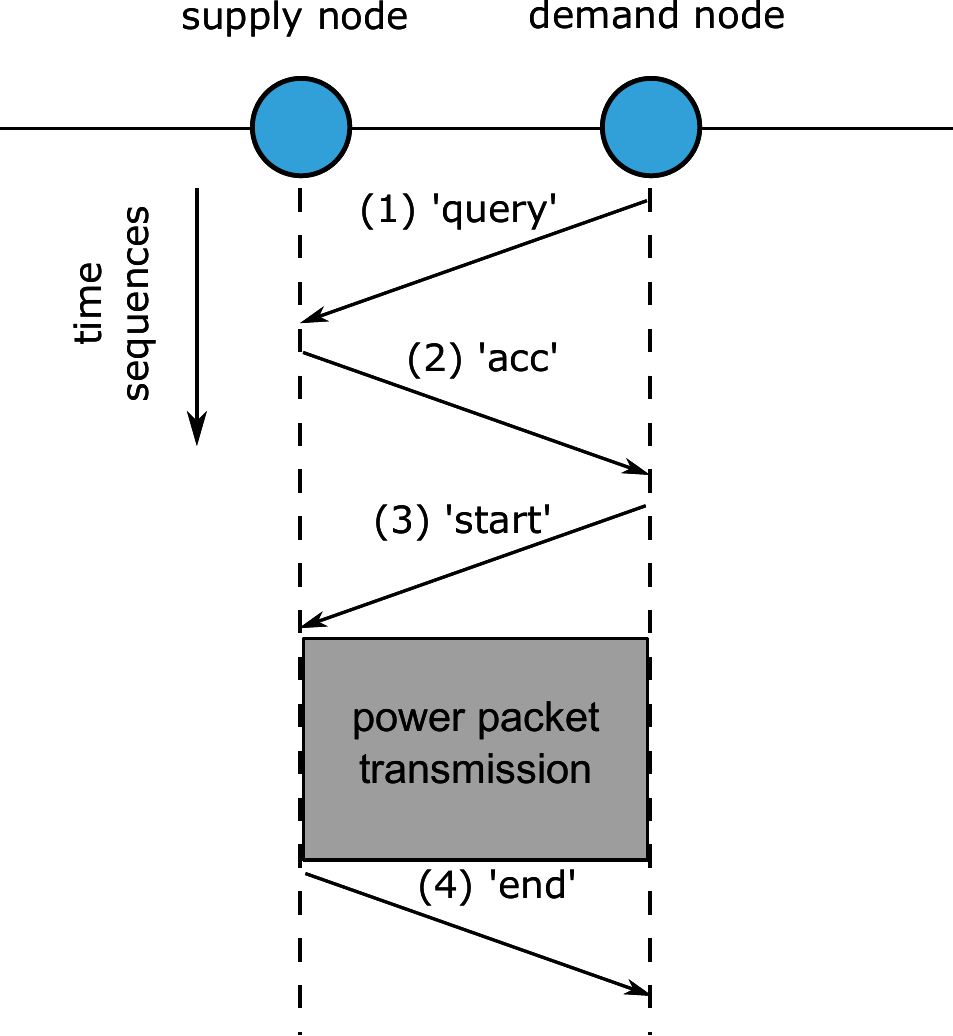}
  \caption{
    A communication logic of the bottom-up method between two power routers. 
    In the bottom-up method, 
    the power packet transmission is triggered by the demand-side node; 
    thus, 
    the communication sequence is started with the ``query'' message from the demand-side node.
    Conversely, 
    in the top-down method,
    the power packet transmission is triggered by the supply-side node,
    which results in the omission of the ``query'' message 
    and the sequence being started with the 'acc' message from the supply-side node.
  }
  \label{f:commun_logic}
\end{figure}

\begin{figure}[t!]
  \centering
  \includegraphics[width=0.9\textwidth]{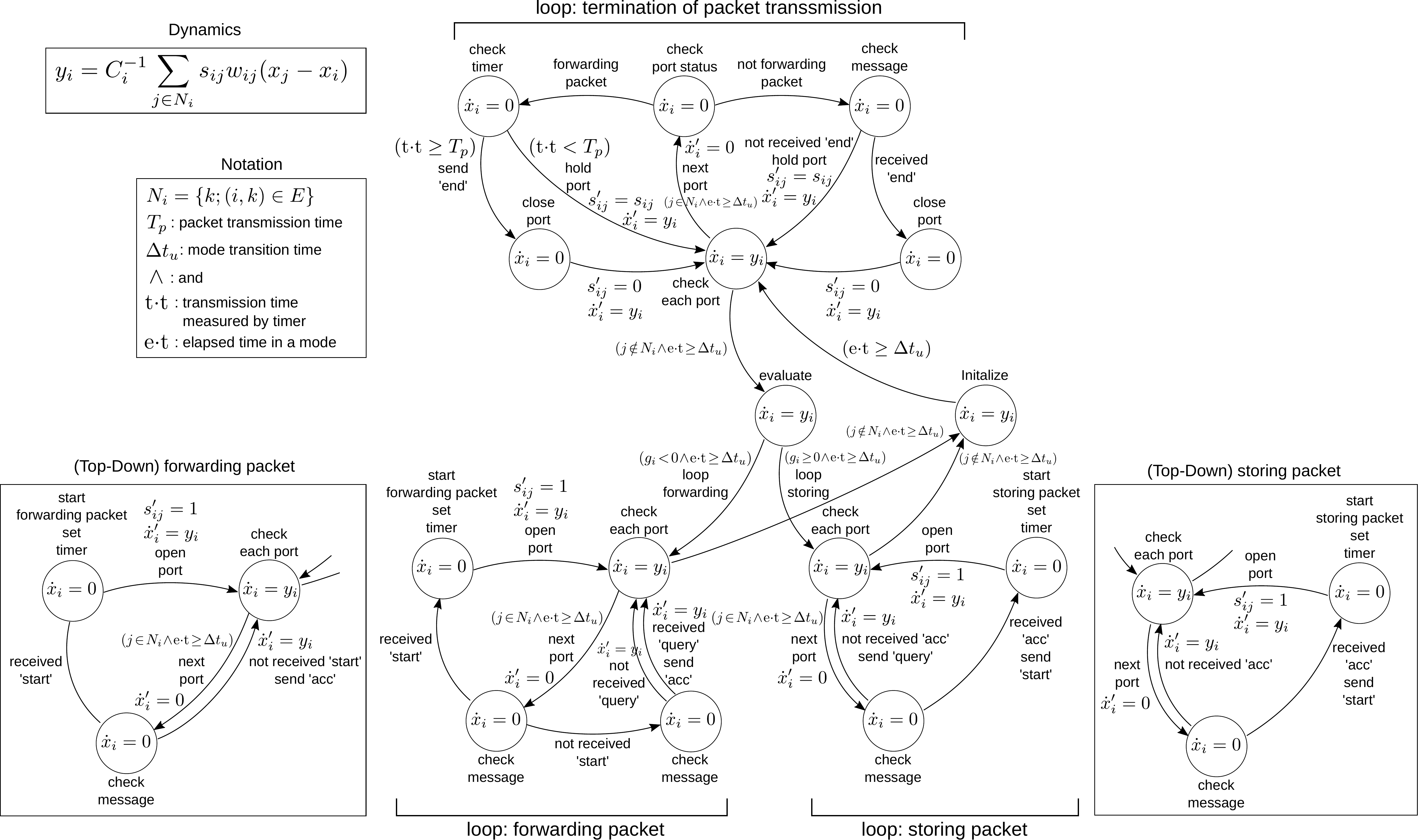}
  \caption{
    Decentralized algorithms of the bottom-up method and top-down method for a power router.
    The algorithm is divided into three loops: 
    termination of packet transmission, 
    packet forwarding, 
    and packet storing.
    Here, the parentheses () indicate the transition conditions, 
    and the superscript-($'$) denotes the updated values when transitioned.
  }
  \label{fig:decentralized_algorithms}
\end{figure}

\figref{f:commun_logic} presents the communication sequences between two adjacent nodes for power packet transmission.
To accomplish the power packet transmissions,
the communication sequences consist of four messages; ``query,'' ``acc,'' ``start,'' and ``end,''
which refer to ``packet-request,'' ``transmission-available,'' ``transmission-start,'' and ``transmission-end,'' respectively.
First, 
if a node satisfies $g_i(t) \ge 0$ given in \eqref{e:eval},
it sends a ``query'' signal to the adjacent nodes.
Then, 
the ``acc'' message is returned if the adjacent node $j$ satisfies the condition $g_j(t) < 0$.
\begin{equation}
  g_i(t) = C_i^{-1}w_{ij^*}(v_{j^*} - v_i) ~~ {\rm s.t.} ~~ j^* = \argmax_{j\in\{k;(i,k) \in E\}}w_{ij}|v_j-v_i|
  \label{e:eval}
\end{equation}
Note that $g_i$ is given as a potential consensus-based distribution model in \eqref{e:condyn_ct_agent},
and also indicates the maximum absolute value of the voltage difference between node $i$ and the adjacent node $j$.
After receiving the ``acc'' message,
the demanding node sends the ``start'' message and prepares to receive a power packet.
When the ``start'' message is received,
the adjacent node sends a power packet to the target node,
which results in a routing $w(e_{ij}) = r_{ij}^{-1}$ between two nodes during single power packet transmission.
After the termination of the transmission,
the supply node sends the ``end'' message to the demand node,
and the path $e_{ij}$ is unrouted with $w(e_{ij}) = 0$.

Based on the above schematics,
a decentralized algorithm is provided in \figref{fig:decentralized_algorithms}.
The algorithm is divided into three loops: 
``termination of transmission,'' ``forwarding packet,'' and ``storing packet''.
The decentralized algorithm begins with the ``initialize'' mode.
Then, 
when a node satisfies a condition ${\rm e \!\cdot\! t} \ge \Delta t_u$ 
meaning that the duration of the mode exceeds the given value $\Delta t_u$, 
it transitions to the ``termination of transmission'' loop.
In this loop,
the node checks whether it has received the ``end'' message at each port.
If the ``end'' message is received at the specific port,
the node terminates the power packet transmission,
thus eventually closing the corresponding route.
Otherwise, 
the node holds its routing state.
When all ports are checked in the ``termination of transmission'' loop,
the node transitions to the ``evaluate'' mode,
which determines whether it stores or forwards a power packet based on the evaluation function $g_i$.

In the ``storing packet'' loop,
the node checks whether it has received the ``acc'' message from each port.
If the node has received the ``acc'' message, 
it sends back the ``start'' message and initiates the power packet transmission.
Otherwise, 
the node sends the ``query'' message to its adjacent nodes.
After checking all ports,
the node transitions to the ``initialize'' mode.

In the ``forwarding packet'' loop,
the node checks whether it has received the ``query'' message at each port.
If the node has received the ``query'' message, 
it replies the ``acc'' message and waits for the ``start'' message.
After receiving the ``start'' message from the adjacent node,
the node starts to send a power packet.
When a single power packet is completely sent,
the node tags the ``end'' message.
After checking all ports, 
the node transitions to the ``initialize'' mode.

\section{Simulations} \label{sec:sim}
Using the decentralized algorithms, 
two PPN structures are simulated 
to evaluate the feasibility and features of the proposed control method.
It should be noted that the proposed control method is structure-independent,
indicating that the applied algorithms are identical, 
regardless of the network structure.

\subsection{Simulation Settings}
\begin{figure}[t!]
  \centering
  \subfloat[\label{f:nwk_5x1_chain}]{ 
    \includegraphics[width=0.40\textwidth]{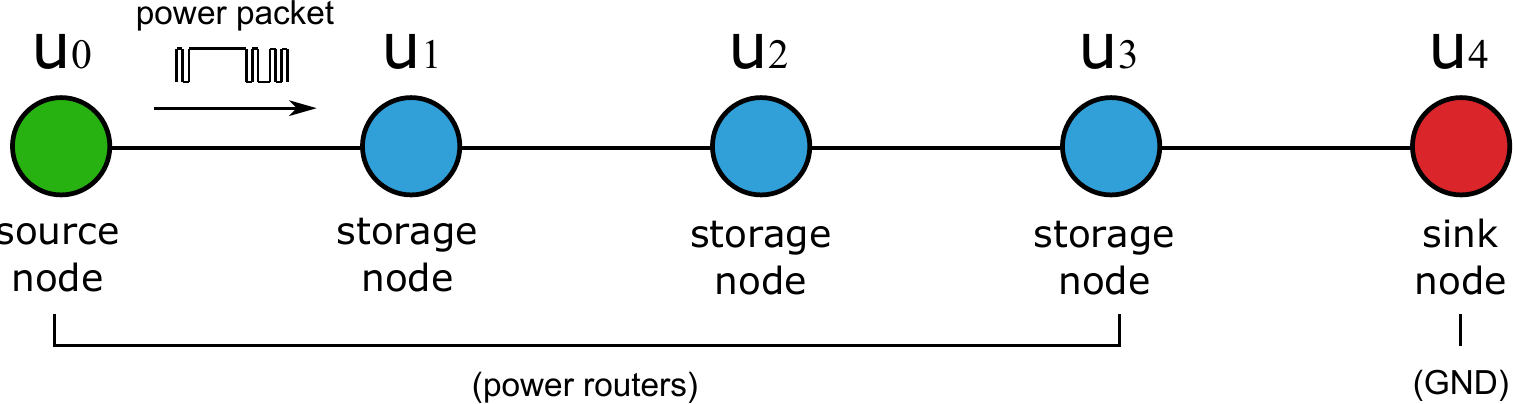}
  }
  \qquad
  \subfloat[\label{f:nwk_tri_mesh}]{ 
    \includegraphics[width=0.40\textwidth]{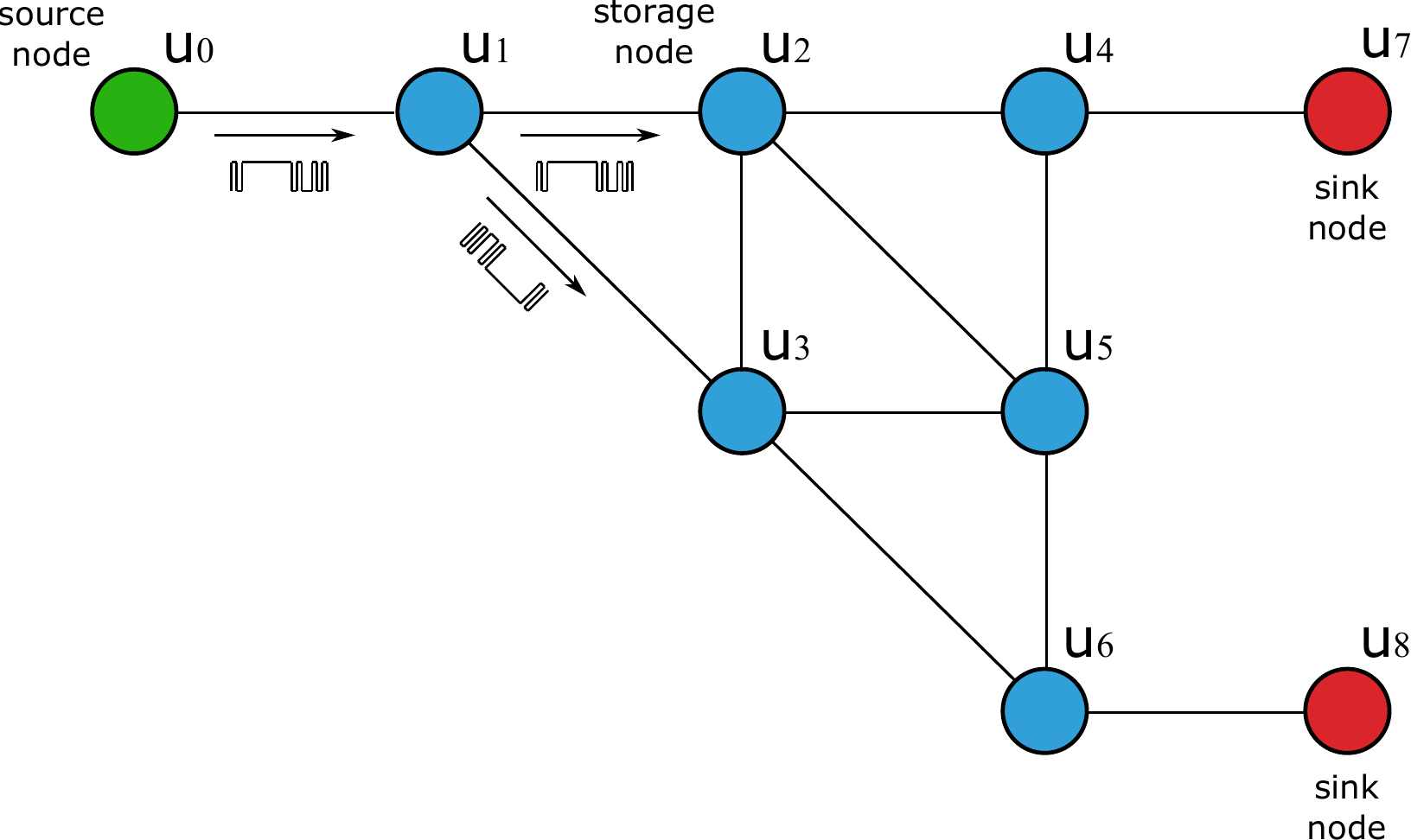}
  }
  \caption{
    Topological structures of a PPN of (a) a 1D chain network and (b) a triangular mesh network. 
    The green nodes indicate power routers with source; 
    the blue nodes, power routers with a storage;
    and the red nodes, sink nodes (GNDs).
  }
  \label{fig:simulated_network}
\end{figure}
To simulate the power packet transmission, 
we assume two networks, 
which are given as a 1D chain structure (\figref{f:nwk_5x1_chain}) and triangular mesh structure (\figref{f:nwk_tri_mesh}).
In each case,
the power packets are generated from the source nodes (indicated in green),
delivered through the storage nodes (indicated in blue),
and then dissipate at sink nodes (indicated in red).
As discussed in Section \ref{subsec:graphical_analysis_ppn},
all the impedances, 
including the switching resistances and loads, 
are considered in the edge weights,
specifically, 
$w_{ij} = r_{ij}^{-1}$ in the case of the connected path $e_{ij}$.

Throughout the simulations presented in \figref{fig:simulated_network},
the following parameters are commonly assumed.
In both cases, 
the voltage of the source node $v_0$ is set to $10\,{\rm V}$,
and the initial voltage of the storage nodes is selected 
from the random values in $[0, 10]$, 
following a uniform distribution.
The capacitance of each storage node is set to $1000\,\mu{\rm F}$.
For the property of the power packet, 
the bit time is set to $3.125\,\mu{\rm s}$,
and the bit length to 100; 
thus, 
the unit duration of the power packet becomes $T_p = 312.5\,\mu{\rm s}$.
Moreover, 
10 bit of the unit power packet is considered as an abstract information tag.
Hence, 
during the power packet transmission of, 
power flow is allowed in only $90\,\%$ of its duration.
For the power routers corresponding to the source nodes and storage nodes,
the switching resistance is set to $1\,\Omega$.
This leads to $0.5\,\Omega^{-1}=(1\,\Omega+1\,\Omega)^{-1}$ of the edge weight when connected,
because two switching devices are located on the paths of the source node and storage node.
In the case of a disconnected condition, 
the edge weight is set to $0$.
In addition, 
we assume $50\,\Omega$ of loads adjacent to sink nodes,
which gives a storage-to-sink edge weight of $0.0196\,\Omega^{-1}=(1\,\Omega+50\,\Omega)^{-1}$.
For simplicity,
we assume that the paths of the storage node and sink node are connected during the entire simulation.
In the applied decentralized algorithms, 
the mode transition time is set to $\Delta t_u = 10\,\mu{\rm s}$.

In both the simulations presented in \figref{fig:simulated_network},
the bottom-up method and top-down method presented in \figref{fig:decentralized_algorithms} are applied.
In addition to the simulations presented in \figref{f:nwk_tri_mesh}, 
a mixed control method is applied to test the difference between the two proposed control methods.
Specifically,
the top-down method is applied to node $u_0$, $u_1$, $u_2$, and $u_4$,
whereas the bottom-up method is applied to node $u_3$, $u_5$, and $u_6$.
The dynamics of power is calculated using \eqref{e:power_dynamics}.
Given these simulation settings, 
a power packet transmission is simulated 10 times 
for each control method on each network structure presented in \figref{f:nwk_5x1_chain} and \figref{f:nwk_tri_mesh}.

\subsection{Results}
\begin{figure}[t!]
  \centering
  \subfloat[\label{f:results_1d_nwk_bu_time_x_voltage}]{ 
    \includegraphics[width=0.33\textwidth]{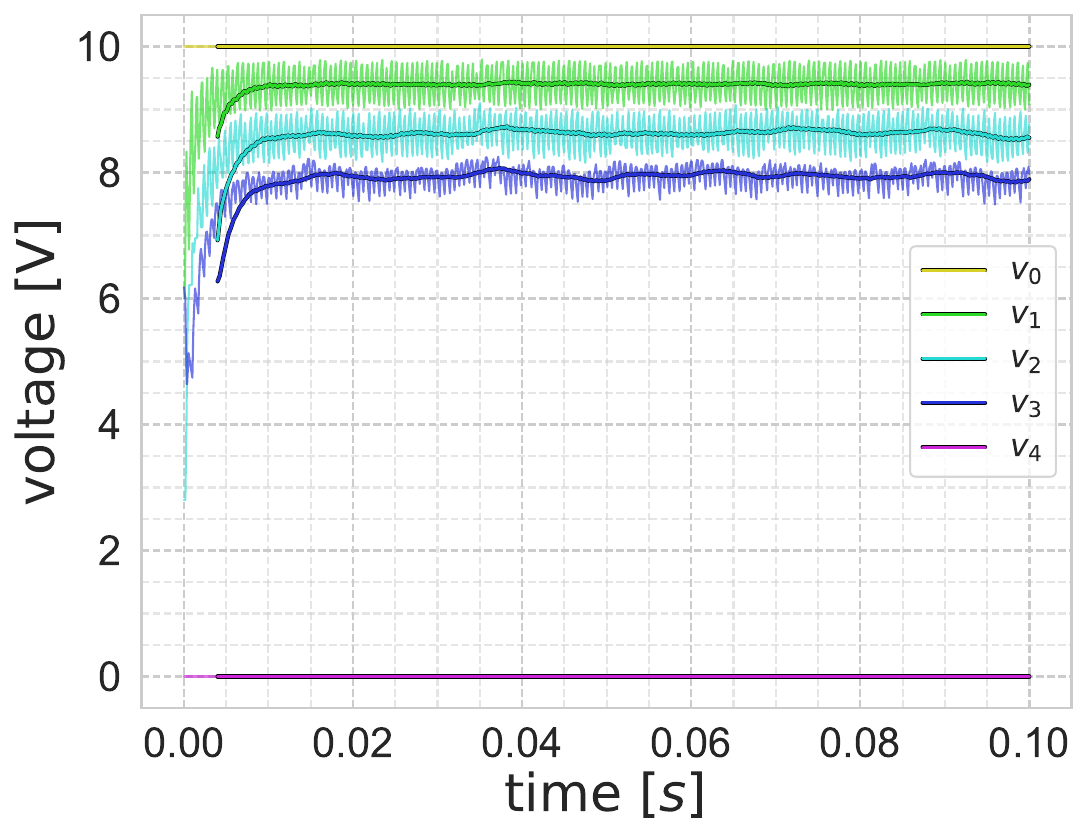}
  }
  \subfloat[\label{f:results_1d_nwk_td_time_x_voltage}]{ 
    \includegraphics[width=0.33\textwidth]{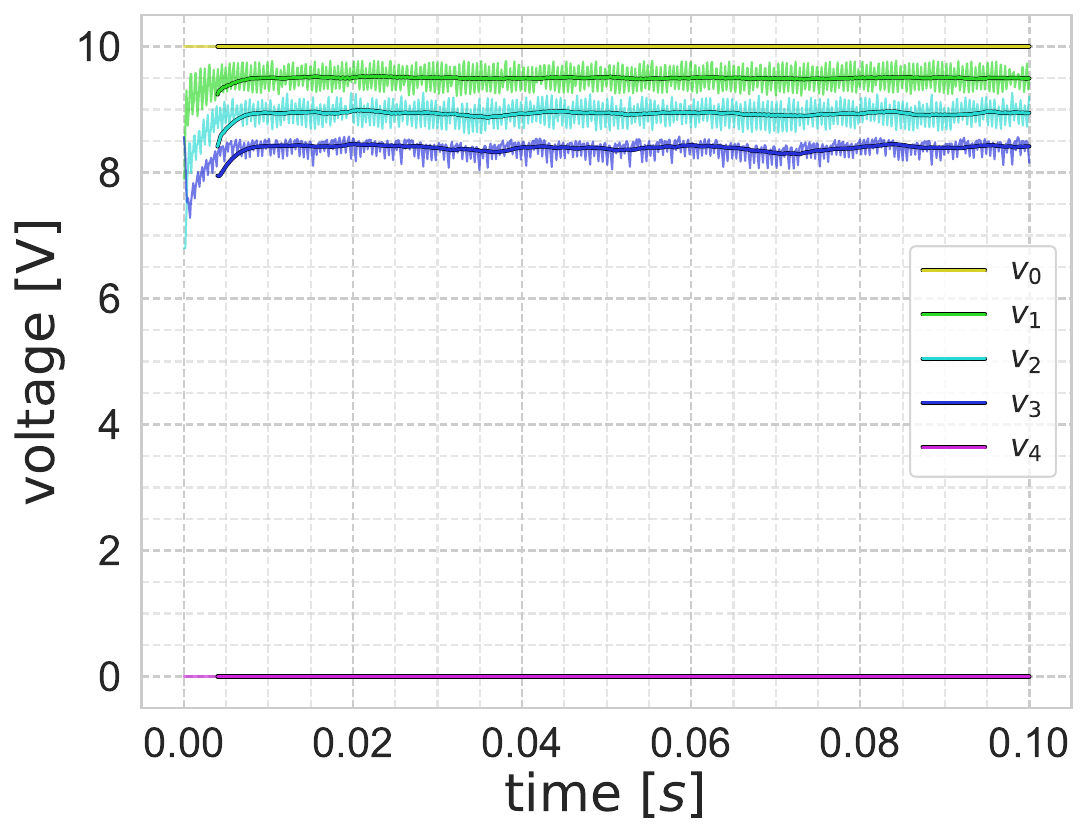}
  }
  \\
  \subfloat[\label{f:results_1d_nwk_bu_time_x_power}]{ 
    \includegraphics[width=0.33\textwidth]{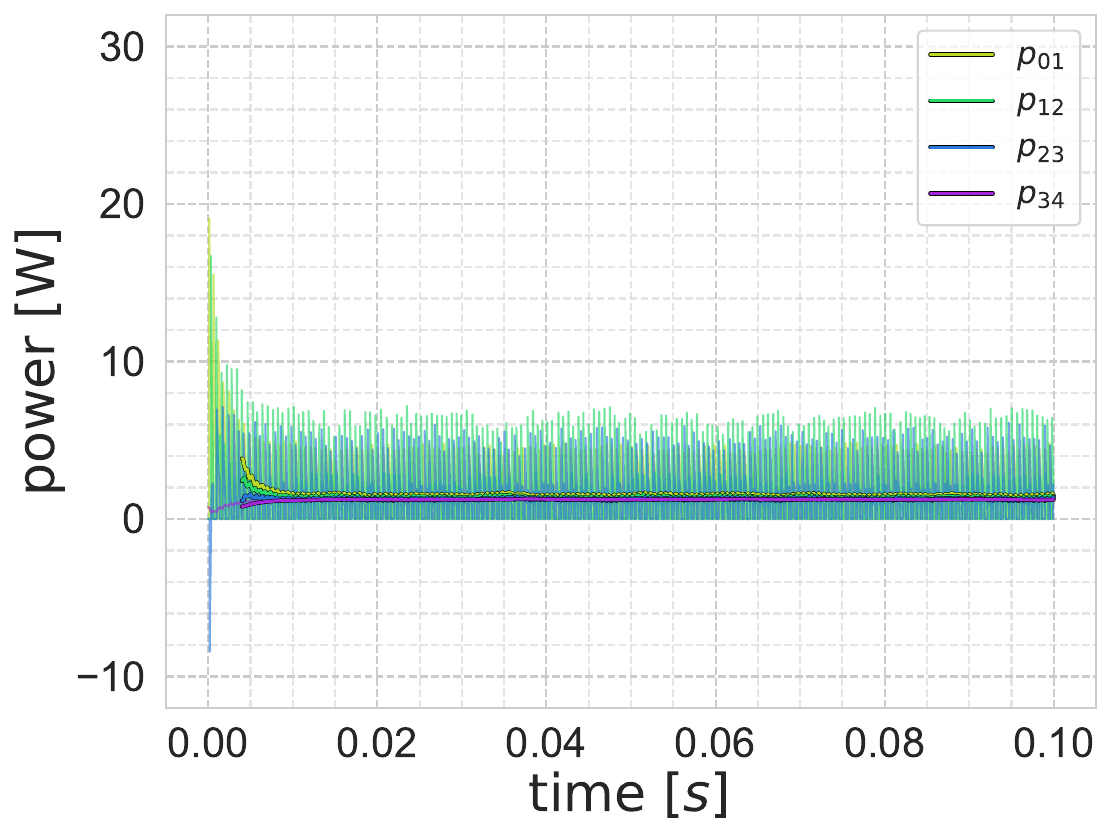}
  }
  \subfloat[\label{f:results_1d_nwk_td_time_x_power}]{ 
    \includegraphics[width=0.33\textwidth]{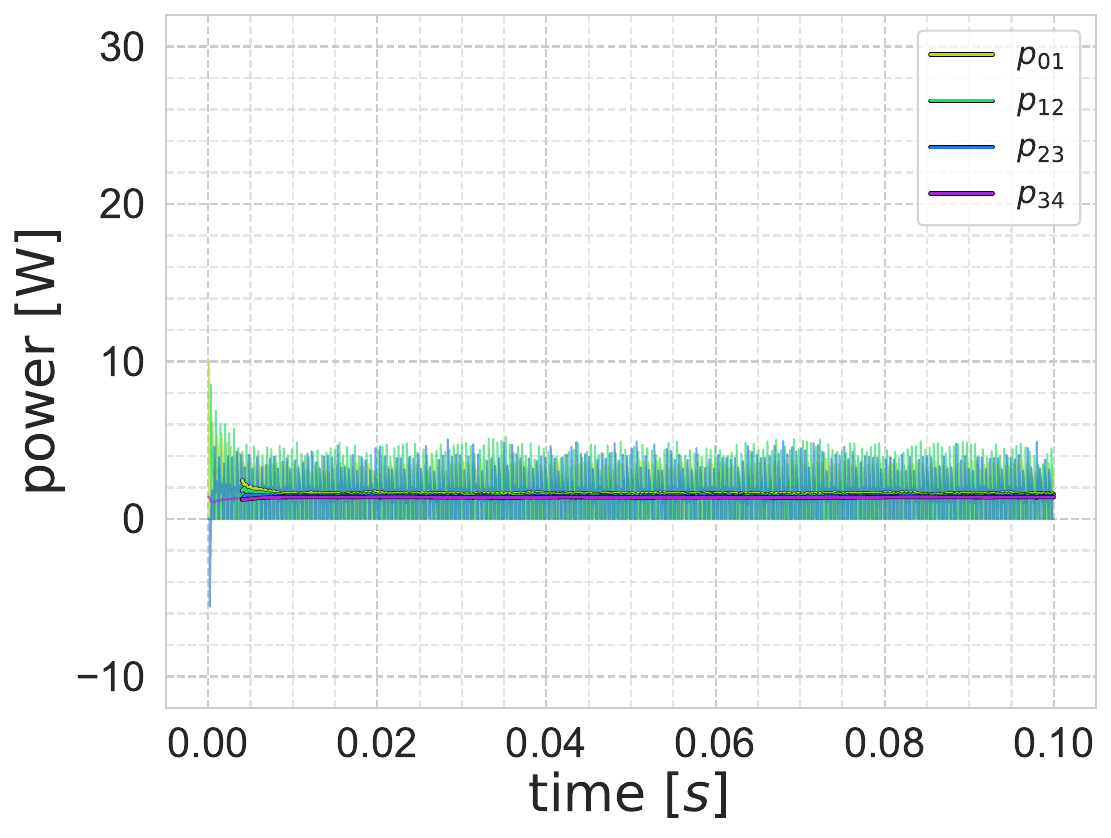}
  }
  \caption{
    A simulation case of the time series data of the node voltages and path powers on 1D chain network (\figref{f:nwk_5x1_chain}). 
    From the top left, 
    each subfigure represents 
    the voltage data of the (a) bottom-up method and (b) top-down method
    as well as the power data of the (c) bottom-up method and (d) top-down method,
    respectively.
    The solid lines indicate the moving average with a window size of $1.25\,{\rm ms}$,
    which is based on the light-colored actual data.
  }
  \label{fig:results_1d_nwk_timeseries}
\end{figure}

\begin{figure}[t!]
  \centering
  \subfloat[\label{f:results_1d_nwk_bu_node_x_voltage}]{ 
    \includegraphics[width=0.33\textwidth]{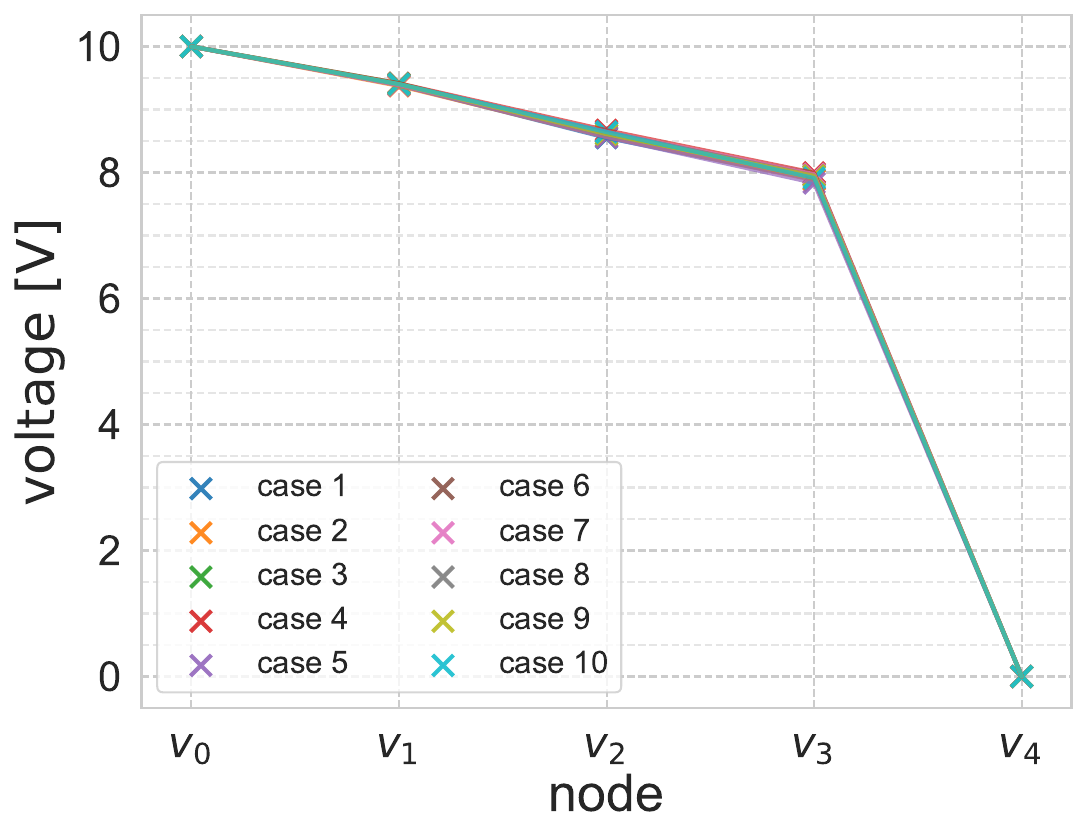}
  }
  \subfloat[\label{f:results_1d_nwk_td_node_x_voltage}]{ 
    \includegraphics[width=0.33\textwidth]{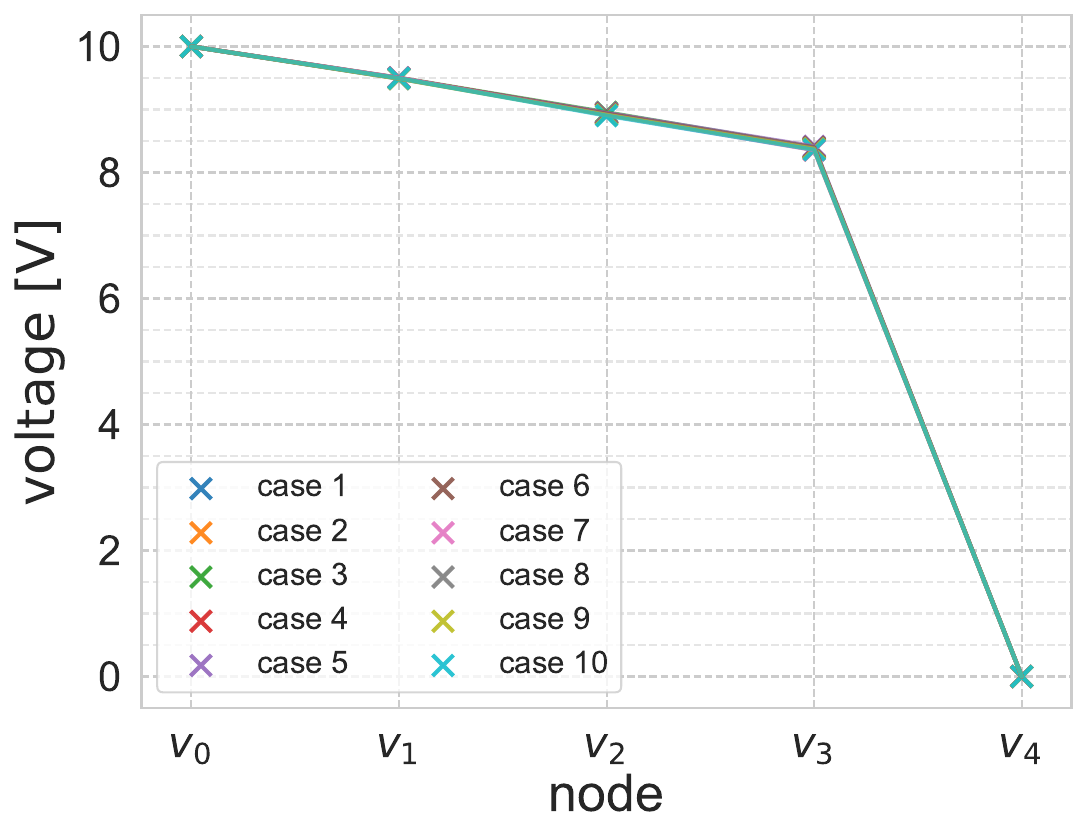}
  }
  \\
  \subfloat[\label{f:results_1d_nwk_bu_path_x_power}]{ 
    \includegraphics[width=0.33\textwidth]{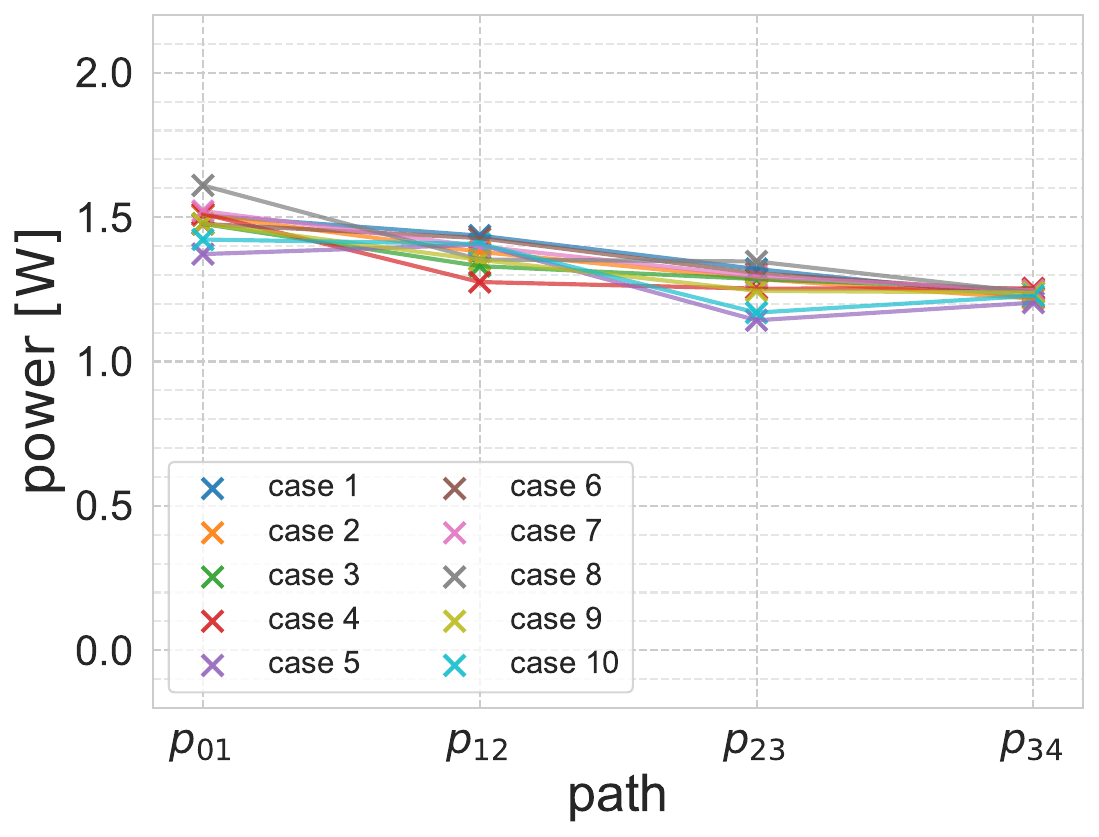}
  }
  \subfloat[\label{f:results_1d_nwk_td_path_x_power}]{ 
    \includegraphics[width=0.33\textwidth]{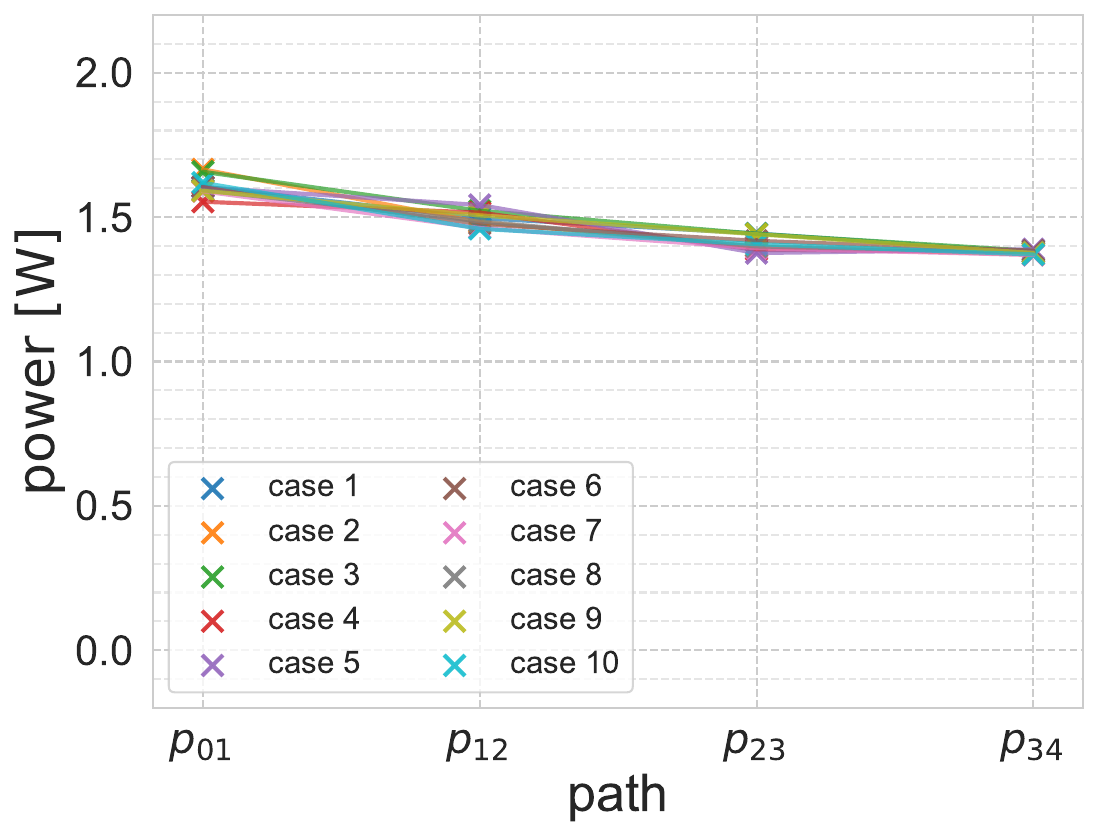}
  }
  \caption{The voltage distribution and power throughputs of the moving average data on 1D chain network (\figref{f:nwk_5x1_chain}) at $t=0.1\,{\rm s}$.
  From the top left,  
  each subfigure represents 
  the voltage data of the (a) bottom-up method and (b) top-down method,
  as well as the power data of the (c) bottom-up method and (d) top-down method,
  respectively.}
  \label{fig:results_1d_nwk_endpoint}
\end{figure}

For the given PPN in \figref{fig:simulated_network},
power packet transmissions were simulated using the bottom-up, top-down, and mixed control methods. 
Each simulation recorded the time series for node voltages and path power flows 
as well as the distribution of voltage and power at the end time of the simulations.
For the results of the 1D chain network presented in \figref{f:nwk_5x1_chain},
a simulation case with time series for node voltages and path powers was obtained, 
as in \figref{fig:results_1d_nwk_timeseries};
each of the end points for 10 cases was obtained as in \figref{fig:results_1d_nwk_endpoint}.

From the results presented in \figref{f:results_1d_nwk_bu_time_x_voltage} and \figref{f:results_1d_nwk_td_time_x_voltage},
the bottom-up method exhibited more fluctuations than the top-down method.
This is because the bottom-up method requires extra communication sequences starting with the ``query'' message
compared with the top-down method.
Accordingly, 
this result indicates that the rate of the power packet transmission using the bottom-up method is slightly low,
which resulted in the low voltage distribution in nodes 1, 2, and 3, 
as presented in \figref{f:results_1d_nwk_bu_node_x_voltage} 
in comparison with \figref{f:results_1d_nwk_td_node_x_voltage}.
This also caused higher differences in the voltage distributions among nodes 0, 1, 2, and 3 
using the bottom-up method.
As a result, 
larger fluctuations in the path of power flows were observed,
as presented in \figref{f:results_1d_nwk_bu_time_x_power} 
compared with the result of \figref{f:results_1d_nwk_td_time_x_power}.
Moreover, 
in the multiple simulation cases presented in \figref{f:results_1d_nwk_bu_path_x_power} and \figref{f:results_1d_nwk_td_path_x_power}, 
a variance of formed power throughputs is higher
using the bottom-up method compared with the top-down method.
These results indicate that the power packet transmission is more sparse 
in the case of transmission among nodes using the bottom-up method
compared with that using the top-down method.

\begin{figure}[t!]
  \centering
  \subfloat[\label{f:results_tri_nwk_bu_time_x_voltage}]{ 
    \includegraphics[width=0.33\textwidth]{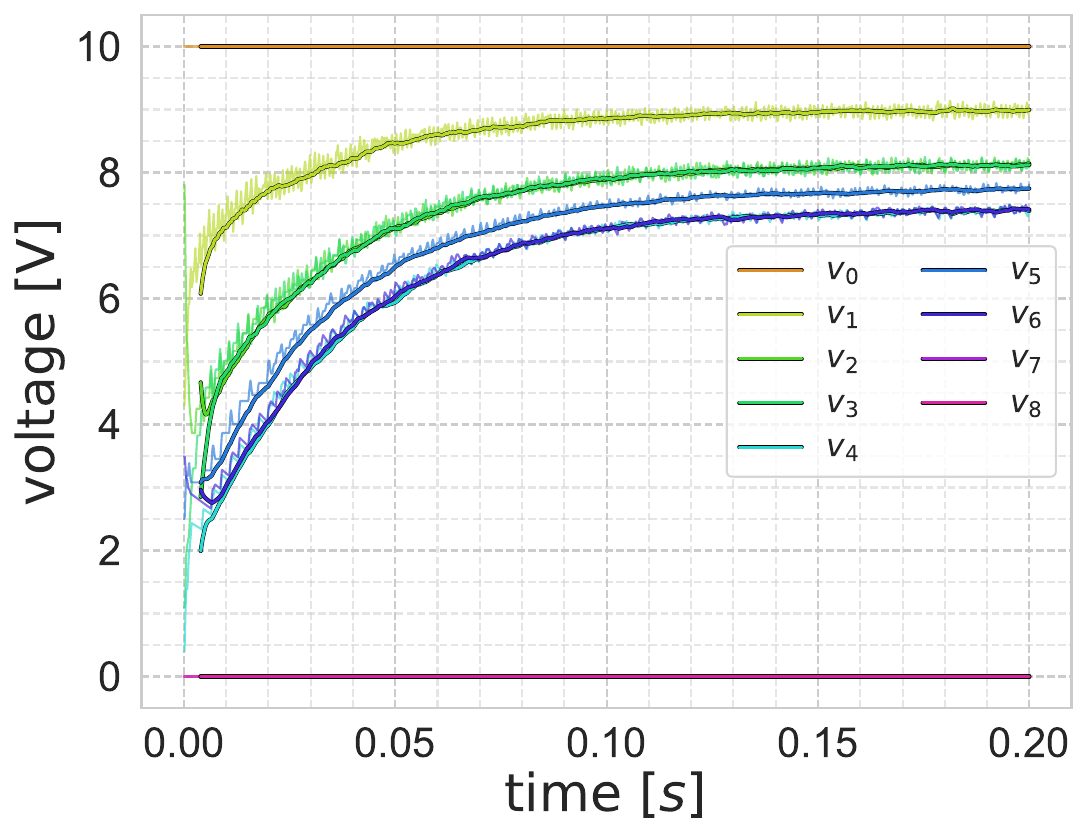}
  }
  \subfloat[\label{f:results_tri_nwk_td_time_x_voltage}]{ 
    \includegraphics[width=0.33\textwidth]{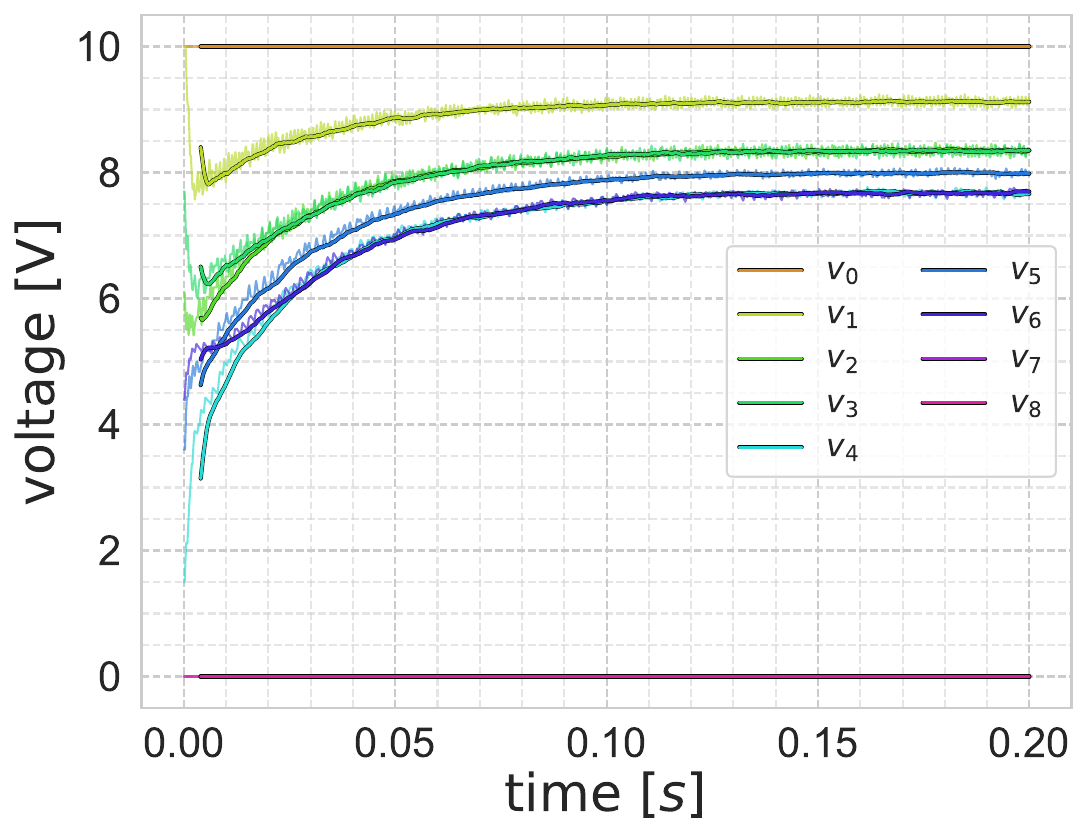}
  }
  \subfloat[\label{f:results_tri_nwk_mixed_time_x_voltage}]{ 
    \includegraphics[width=0.33\textwidth]{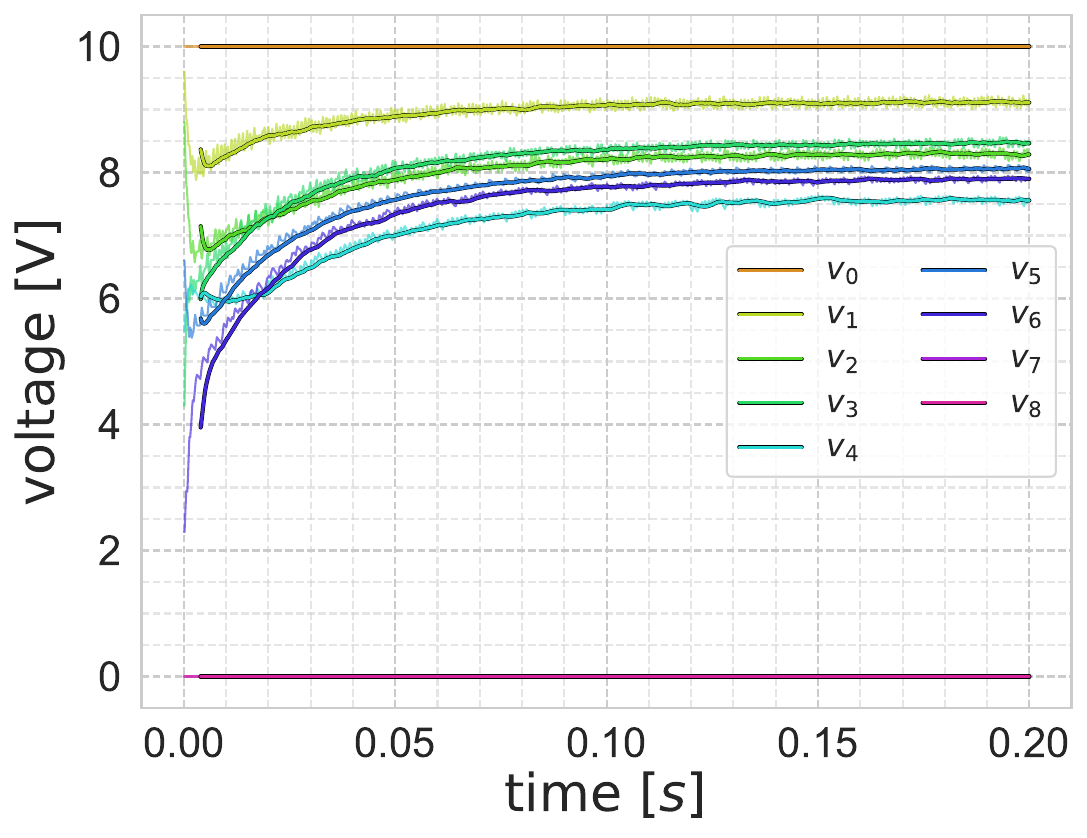}
  }
  \\
  \subfloat[\label{f:results_tri_nwk_bu_time_x_power}]{ 
    \includegraphics[width=0.33\textwidth]{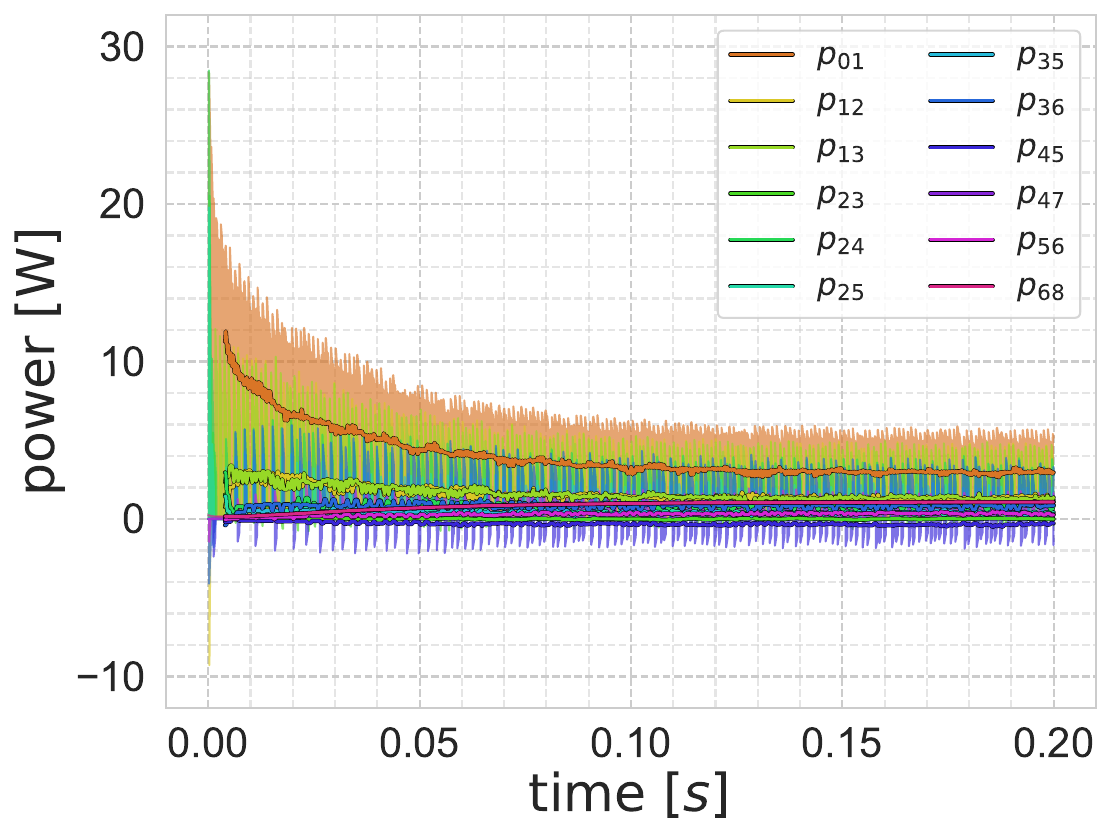}
  }
  \subfloat[\label{f:results_tri_nwk_td_time_x_power}]{ 
    \includegraphics[width=0.33\textwidth]{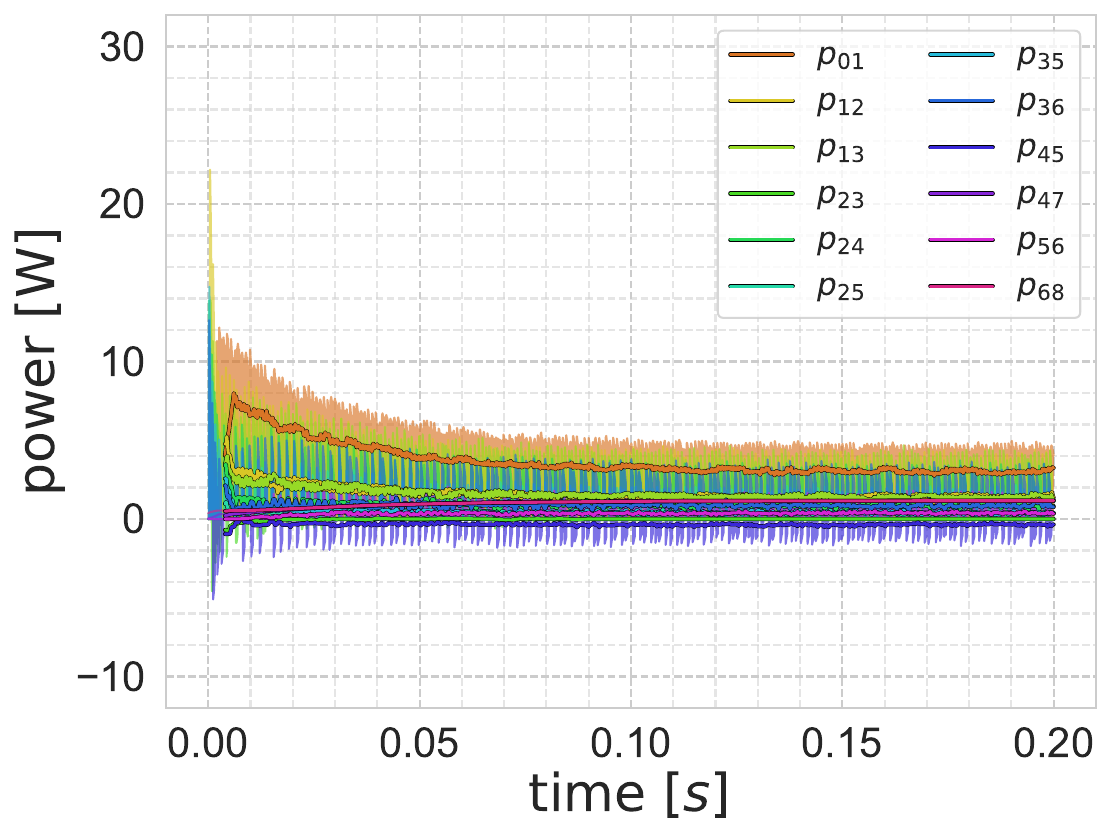}
  }
  \subfloat[\label{f:results_tri_nwk_mixed_time_x_power}]{ 
    \includegraphics[width=0.33\textwidth]{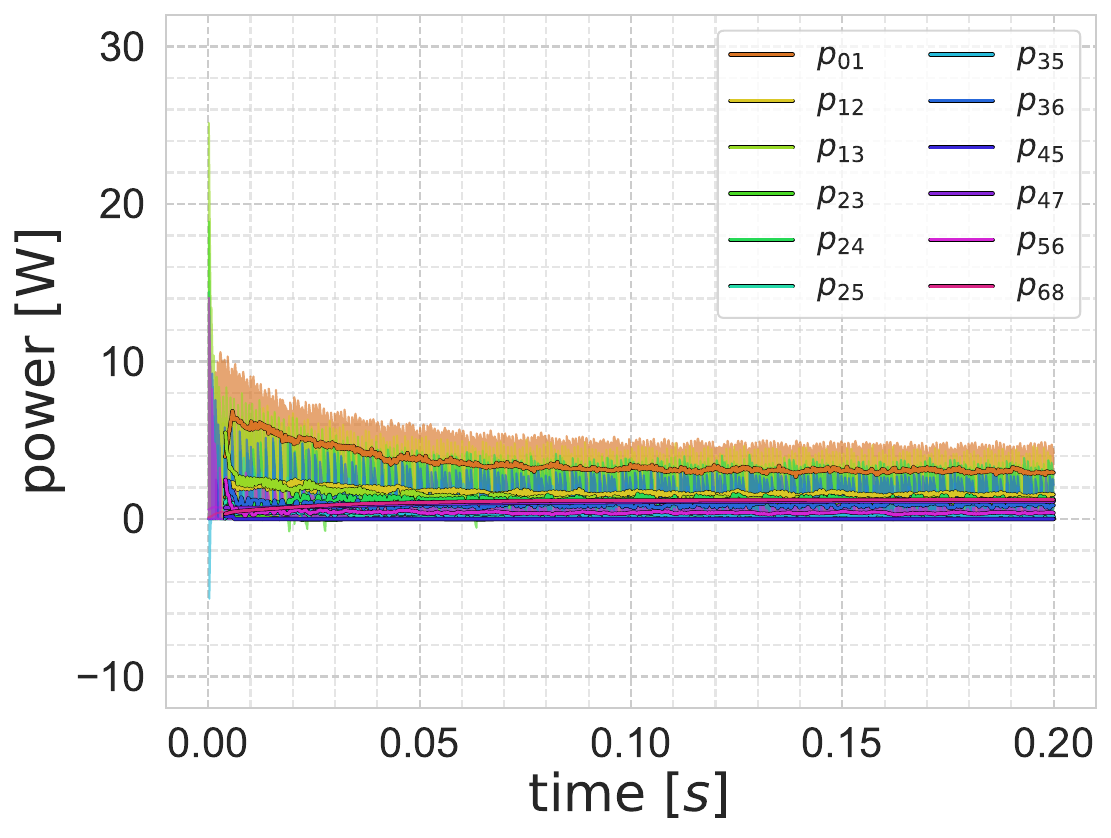}
  }
  \caption{A simulation case of time series data of node voltages and path powers over time on triangular mesh network (\figref{f:nwk_tri_mesh}). 
    From the top left, 
    each subfigure represents 
    the voltage data of the (a) bottom-up method, (b) top-down method, and (c) mixed control method
    as well as the power data of the (d) bottom-up method, (e) top-down method, and (f) mixed control method,
    respectively.
    The solid lines indicate the moving average with a window size of $1.25\,{\rm ms}$,
    which is based on the light-colored actual data.
    }
  \label{fig:results_tri_nwk_timeseries}
\end{figure}

\begin{figure}[t!]
  \centering
  \subfloat[\label{f:results_tri_nwk_bu_node_x_voltage}]{ 
    \includegraphics[width=0.33\textwidth]{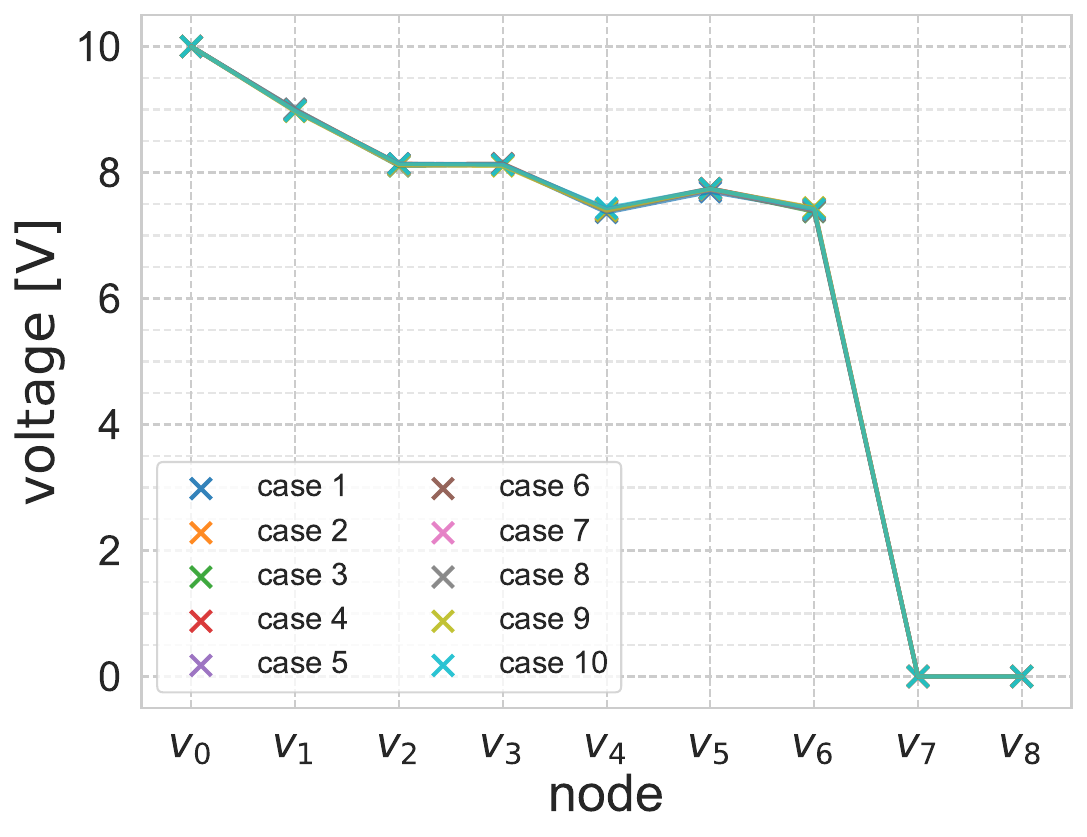}
  }
  \subfloat[\label{f:results_tri_nwk_td_node_x_voltage}]{ 
    \includegraphics[width=0.33\textwidth]{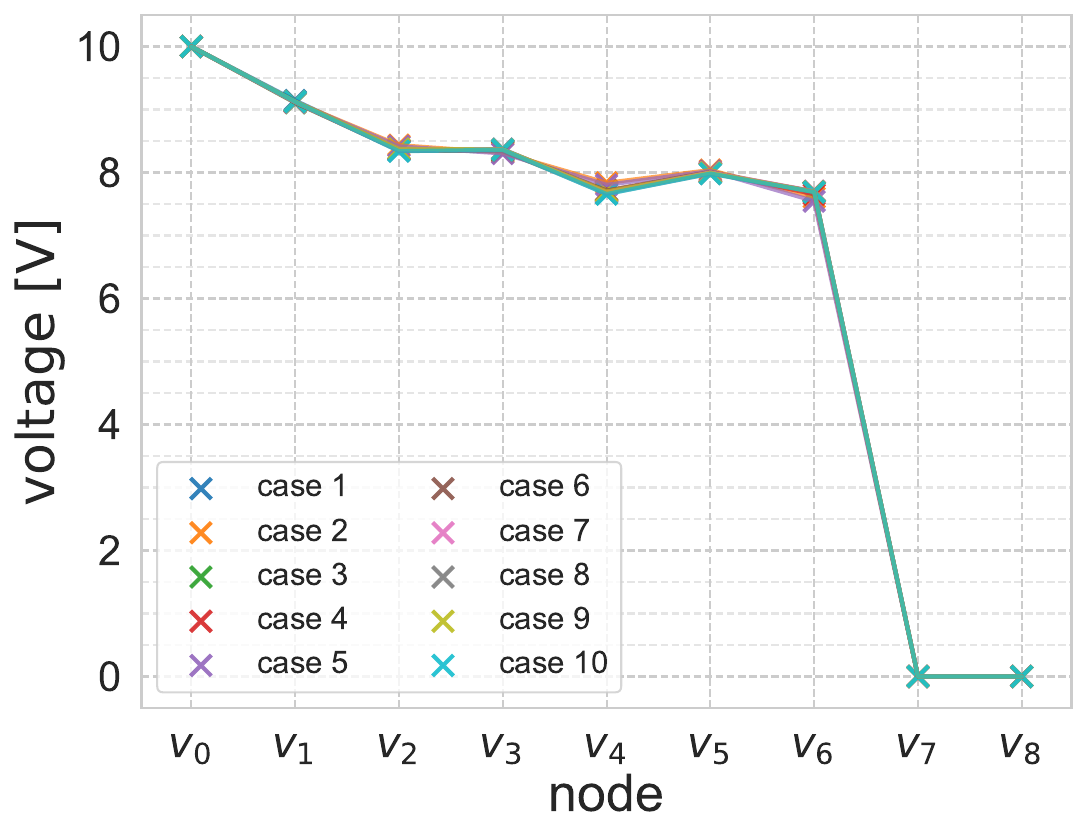}
  }
  \subfloat[\label{f:results_tri_nwk_mixed_node_x_voltage}]{ 
    \includegraphics[width=0.33\textwidth]{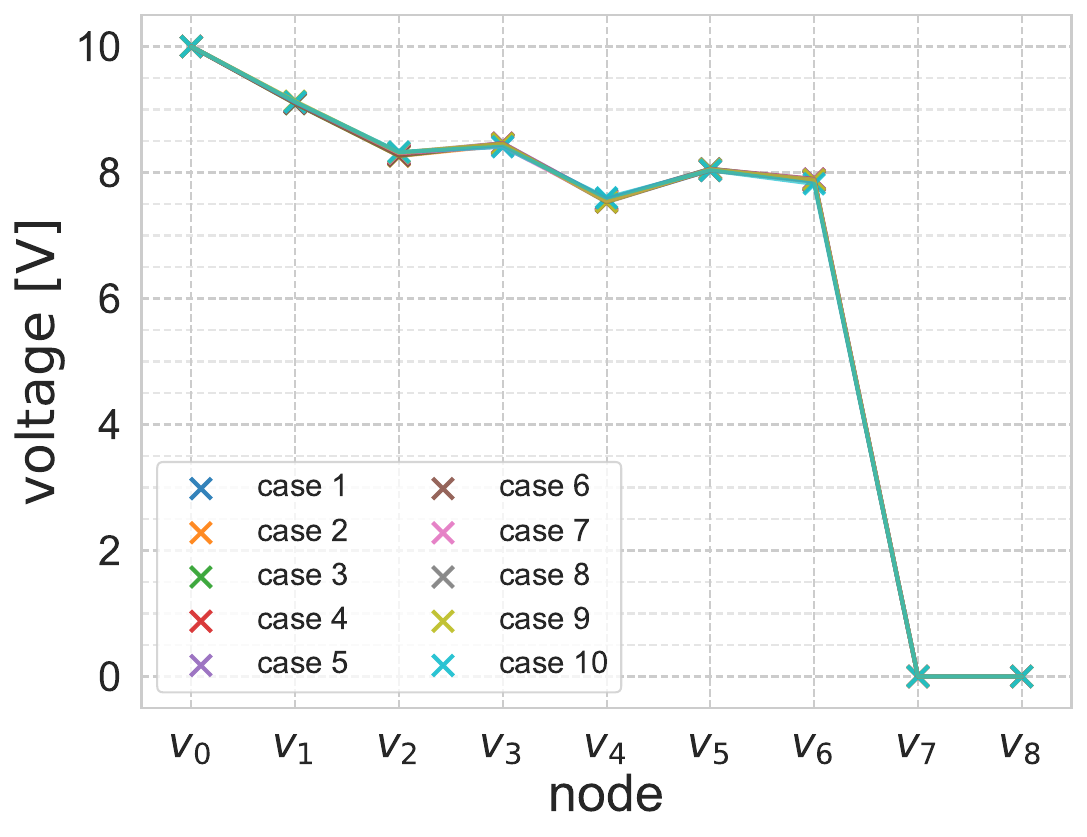}
  }
  \\
  \subfloat[\label{f:results_tri_nwk_bu_path_x_power}]{ 
    \includegraphics[width=0.33\textwidth]{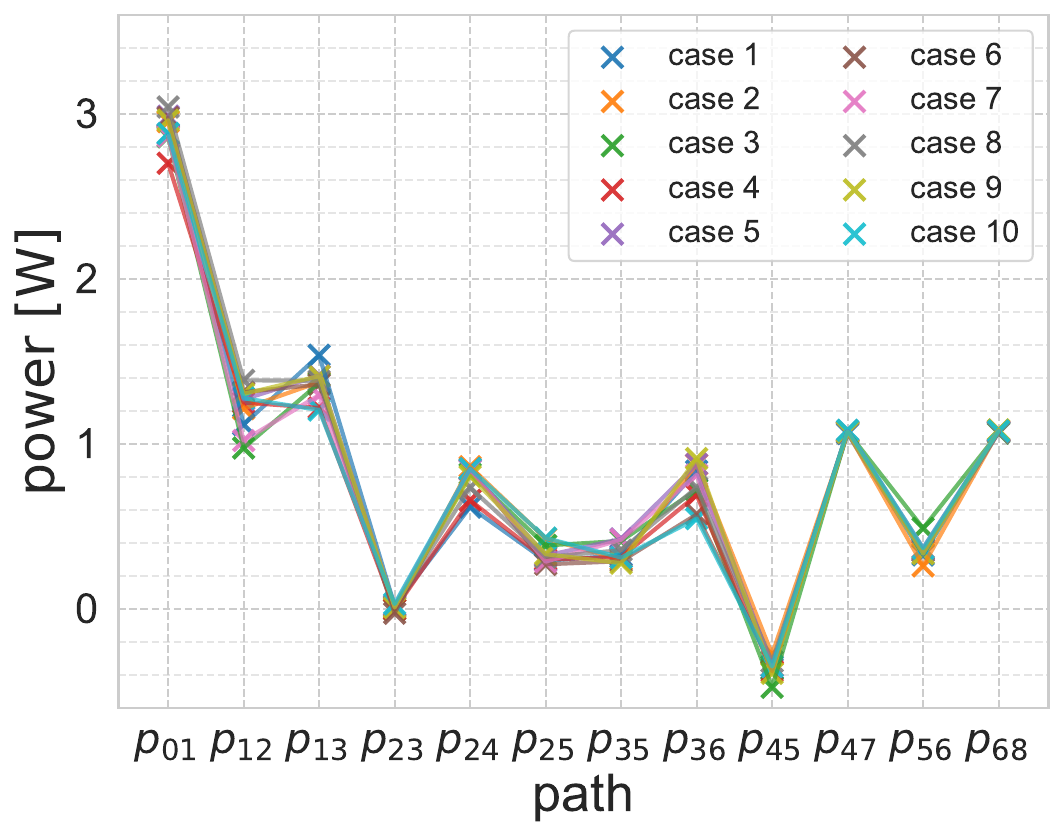}
  }
  \subfloat[\label{f:results_tri_nwk_td_path_x_power}]{ 
    \includegraphics[width=0.33\textwidth]{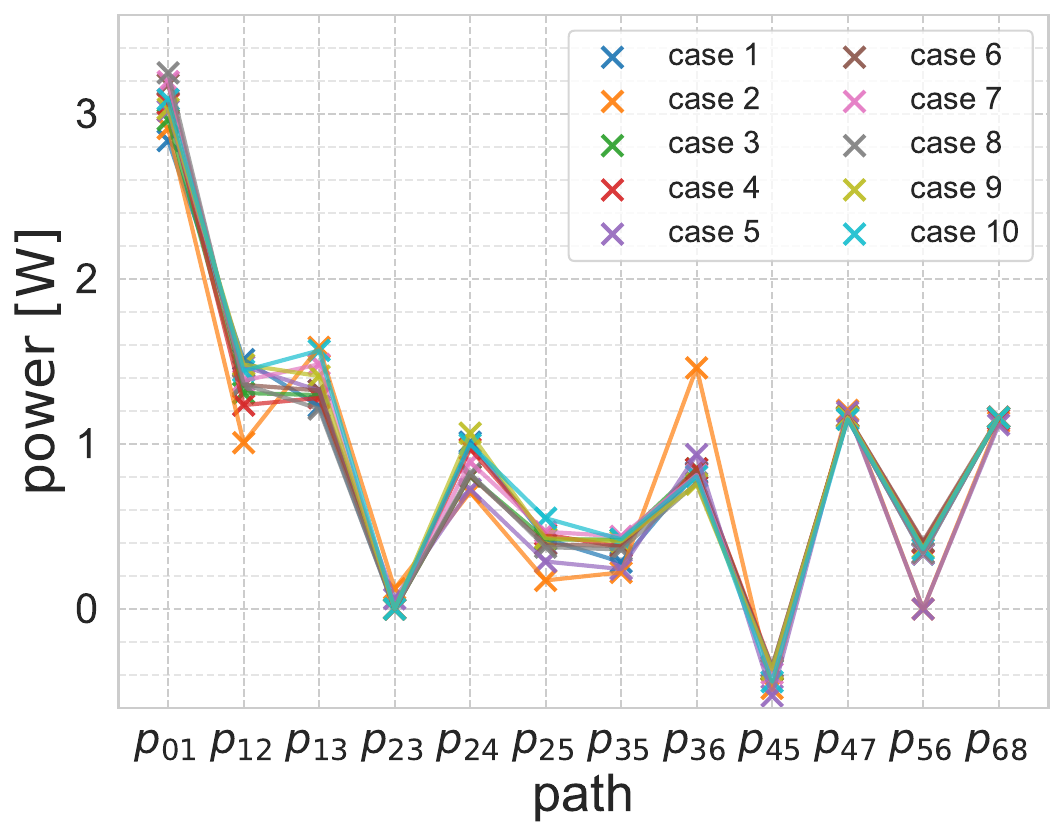}
  }
  \subfloat[\label{f:results_tri_nwk_mixed_path_x_power}]{ 
    \includegraphics[width=0.33\textwidth]{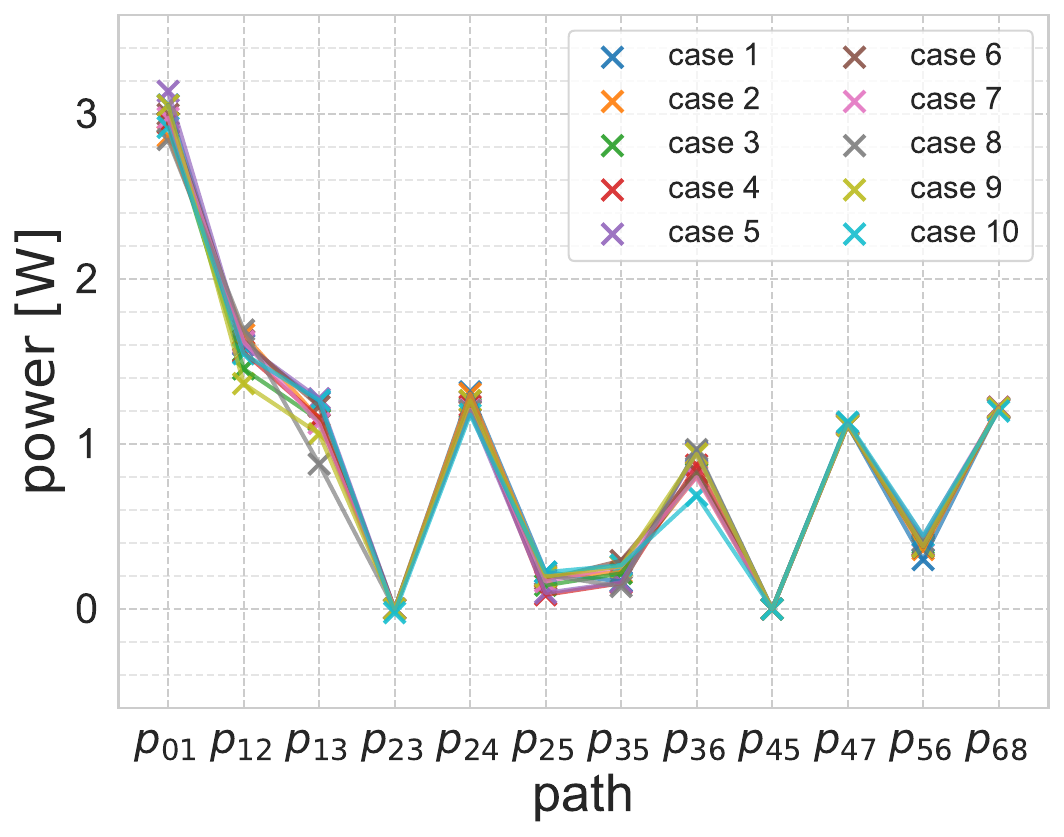}
  }
  \caption{The voltage distribution and power throughputs of moving average data on triangular mesh network (\figref{f:nwk_tri_mesh}) at $t=0.2\,{\rm s}$.
  From the top left,  
  each subfigure represents 
  the voltage data of the (a) bottom-up method, (b) top-down method, and (c) mixed control method,
  as well as the power data of the (d) bottom-up method, (e) top-down method, and (f) mixed control method,
  respectively.
  } 
  \label{fig:results_tri_nwk_endpoint}
\end{figure}

From the simulations performed on the 1D chain structure presented in \figref{f:nwk_5x1_chain},
a plain characteristic can be derived,
{\it i.e.},
in the given decentralized control, 
the top-down method exhibits better performance in power broadcasting.
While the top-down method focuses on the supply-driven packet distribution,
the bottom-up method deals with the demand-driven packet distribution (also called demand response).
In the simulations of the top-down method and the bottom-up method in this paper, 
all node of PPN is basically assumed to be requesting for power packets.
Due to this setting,
it is difficult to capture the remarkable difference between two methods.
This problem is explained in the following simulation with the mixed control method.

In the simulation of a triangular mesh network in \figref{f:nwk_tri_mesh},
the time series data and distribution of voltages and powers were obtained in \figref{fig:results_tri_nwk_timeseries} and \figref{fig:results_tri_nwk_endpoint}, respectively.
By comparing the bottom-up case to the top-down case,
as presented in \figref{f:results_tri_nwk_bu_time_x_voltage}, \figref{f:results_tri_nwk_td_time_x_voltage}, \figref{f:results_tri_nwk_bu_node_x_voltage}, and \figref{f:results_tri_nwk_td_node_x_voltage},
the overall aspect of the voltage distribution was similar 
to that in the simulation performed on the structure presented in \figref{f:nwk_5x1_chain}.
More specifically,
the voltage distribution is gradually reduced according to the topological distance to the source node.
In addition,
the higher path degree of the inflow led to a high voltage status 
as shown in node $v_5$ comparing to nodes $v_4$ and $v_6$ in the same topological distance to the source node.
The overall voltage distribution in the top-down case was higher than that in the bottom-up case.
This result agrees with those from the previous results of \figref{f:results_1d_nwk_bu_node_x_voltage} and \figref{f:results_1d_nwk_td_node_x_voltage}.

From the results of power flows obtained, 
as presented in \figref{f:results_tri_nwk_bu_time_x_power}, \figref{f:results_tri_nwk_td_time_x_power}, \figref{f:results_tri_nwk_bu_path_x_power}, and \figref{f:results_tri_nwk_td_path_x_power},
larger fluctuations were observed in comparison with the simulations performed on the 1D chain structure.
If a temporal imbalance or fluctuation exists in the same stage's nodes,
resilience is induced 
by the consensus-based distribution model employed in the decentralized algorithms of each node.
Specifically, 
in the voltage fluctuation shown as in \figref{f:results_tri_nwk_bu_time_x_voltage} and \figref{f:results_tri_nwk_td_time_x_voltage},
the voltage distributions of $v_4$ and $v_6$ were supposed to be balanced 
due to the symmetry of the network topology and the same algorithm applied.
However, 
for case 2 and 5 in \figref{f:results_tri_nwk_td_node_x_voltage},
$v_4$ was slightly higher than $v_6$,
thus causing a small gap between $p_{47}$ and $p_{68}$, 
as presented in \figref{f:results_tri_nwk_td_path_x_power}.
These are presumed to be temporary and on the balancing process, 
seeing that $p_{36}$ is larger than $p_{24}$ at the given period,
as well as $p_{12}$ and $p_{13}$.
Thus, 
it can be inferred that the unbalanced status of $v_4$ and $v_6$ can be adjusted
by balancing the power flows on the related paths: $e_{12}$, $e_{13}$, $e_{24}$, and $e_{36}$.
This result indicates that 
the proposed control method is capable of energy management, 
especially power balancing and resilience.

By employing the top-down method for nodes ${u_0, u_1, u_2}$, and $u_4$
and the bottom-up method for node $u_3, u_5$, and $u_6$,
the simulation results of the mixed control method were obtained 
as in \figref{f:results_tri_nwk_mixed_time_x_voltage}, \figref{f:results_tri_nwk_mixed_time_x_power}, \figref{f:results_tri_nwk_mixed_node_x_voltage} and \figref{f:results_tri_nwk_mixed_path_x_power}.
As presented in \figref{f:results_tri_nwk_mixed_time_x_voltage} and \figref{f:results_tri_nwk_mixed_node_x_voltage},
the formed voltage distribution of the bottom-up side was larger than that of the top-down side,
which is figured out by $v_6 > v_4$ and $v_3 > v_2$. 
This result indicates that power packet transmission is prioritized on the bottom-up side.
Unlike the simulations of the bottom-up method or top-down method, 
the packet transmission between nodes 4 and 5 was regulated, 
{\it i.e.}, $p_{45} = 0$.
In the path between two nodes applied with different methods, 
the demand-side node requests power packets from its neighbor nodes,
whereas the supply-side node does not respond to the requests with a constraint logic of the top-down algorithm,
in which the ``acc'' message can be sent to nodes with smaller voltage than itself.
As a result,
the biased distributions were observed in the second stage ($v_2$ and $v_3$) and the third stage ($v_4$ and $v_6$), respectively.

The results of the above simulations indicate that 
a PPN is valid in the packet-based energy management and priority-based power control 
using the proposed control method.
The proposed dynamics model applied to the decentralized algorithms, 
{\it i.e.}, consensus dynamics,
the power packet transmission on a PPN follows the spread distribution.
Moreover,
based on the simulation with the mixed control method,
the packet distribution can be adjusted according to the arrangement of nodes with the bottom-up and the top-down methods.

\section{Conclusion} \label{sec:conclusions}
This paper discussed the power distribution model of a PPN 
and proposed a decentralized control for two operational strategies: the top-down method and bottom-up method.
Based on the three-layered consensus model,
we proposed a model of the agent-to-agent packet distribution of a PPN.
This concept has been employed in the decentralized control 
for the operation of the network to follow the intended distribution model.
Based on the established algorithm, 
we simulated the power packet transmissions on a PPN,
in which the local communication was only allowed between power routers.
The results indicated the structural independence of the proposed control method, 
as well as the novel functions for energy management, 
such as power broadcasting and priority-based power control.
The use of the proposed control methods is expected to improve the scalability of the network topology
and to solve the problem of a small-sized power system requiring delicate energy management 
due to the large and frequent fluctuations in the supply/demand conditions. 

The proposed control method provides the possibility of advanced energy management with diverse applications;
still, 
the system itself is low-observable.
Since the control relies on agent-to-agent local interactions,
an additional sensor system might be necessary to capture the changing dynamical aspects of a PPN.
This is because the distribution model based on consensus dynamics only allows a qualitative analysis;
thus, 
it cannot be directly utilized in the real-time control of a PPN.
Therefore, 
despite the advantages, 
such as scalability, flexibility, and novel functions for energy management,
using the proposed control methods is difficult for handling the global objective of the system,
which is important in applications such as swarm robotics or multi-legged machines.
Considering the possible applications of a PPN,
{\it e.g.}, an energy distribution inspired by biomechanics, 
such interesting questions still remain.

\section*{Acknowledgments}
This research was partially supported by Cross-ministered Strategic Innovation Promotion (SIP) Program from NEDO.









%
%
%

\end{document}